\documentclass[aps,nofootinbib,longbibliography,prd,twocolumn]{revtex4-1}
\usepackage[babel]{csquotes}
\usepackage{graphicx}
\usepackage{amsmath,amssymb}
\usepackage[colorlinks,citecolor=blue,linkcolor=blue,urlcolor=blue]{hyperref}
\usepackage{mathrsfs}
\usepackage{enumerate}
\usepackage{dsfont}
\usepackage{mathrsfs}
\def\be{\begin{equation}}
\def\ee{\end{equation}}

\newcommand{\dd}{ \mathrm{d}}
\def\bra#1{\mathinner{\langle {#1}|}}
\def\ket#1{\mathinner{|{#1}\rangle}}

\newcommand{\nn}{ \mathrm{n}}
\DeclareMathOperator{\e}{e}

\newcommand{\I}{\rm I}
\newcommand{\II}{\rm II}
\newcommand{\III}{\rm III}
\newcommand{\IV}{\rm IV}

\usepackage{verbatim} 

\usepackage{scrextend} 
\usepackage{float}

\usepackage{xcolor}
\usepackage{xspace}
\usepackage{ifthen}

\newcommand{\marios}[1]%
{\ifthenelse{\equal{\showcomments}{true}}%
{{\color{red}{\small \textbf{MC:} #1}}}{\xspace}}%

\newcommand{\ct}[1]%
{\ifthenelse{\equal{\showcomments}{true}}%
{{\color{cyan}{\small \textbf{CT:} #1}}}{\xspace}}%

\newcommand{\showcomments}{true}

\begin{document}

\title{Plank Stars lifetime}

\title{Characteristic Time Scales for the Geometry Transition of a Black Hole to a White Hole from Spinfoams}
\author{Marios Christodoulou}
\thanks{christod.marios@gmail.com}

\affiliation{Institute for Quantum Optics and Quantum Information (IQOQI) Vienna, Austrian Academy of Sciences, Boltzmanngasse 3, A-1090 Vienna, Austria}
\affiliation{\small
\mbox{CPT, Aix-Marseille Universit\'e, Universit\'e de Toulon, CNRS, F-13288 Marseille, France.} }
\author{Fabio D'Ambrosio\thanks{fabio.dambrosio@gmx.ch}}
\affiliation{\small
\mbox{CPT, Aix-Marseille Universit\'e, Universit\'e de Toulon, CNRS, F-13288 Marseille, France.} }
\date{\small\today}

\begin{abstract}
Quantum fluctuations of the metric may provide a decay mechanism for black holes through a transition to a white hole geometry.  
Previous studies formulated Loop Quantum Gravity amplitudes with a view to describe this process. We identify two timescales to be extracted  which we call the crossing time and the lifetime and complete a calculation that gives explicit estimates using the asymptotics of the EPRL model. The crossing time is found to scale linearly in the mass, in agreement with previous results by Ambrus and H\'aj\'i\v{c}ek and more recent results by Barcel\'o, Carballo--Rubio and Garay. The lifetime is found to depend instead on the spread of the quantum state, and thus its dependence on the mass can take a large range of values. This indicates that the truncation/approximation used here is not appropriate to estimate this observable with any certainty.  The simplest choice of a balanced semiclassical state is shown to yield an exponential scaling of the lifetime in the mass squared. Our analysis only considers 2-complexes without bulk faces, a significant limitation. In particular it is not clear how our estimates will be affected under refinements. This work should be understood as a step towards a fuller calculation in the context of covariant Loop Quantum Gravity. 

\end{abstract}

\maketitle 

\section{Introduction} \label{sec:Intro}

 In his renowned 1974 letter ``Black hole explosions?'' \cite{hawking_black_1974} Stephen Hawking shows that quantum theory can significantly affect gravity even in low curvature regions, provided that enough time elapses.  Hawking closes with the comment that he has neglected quantum fluctuations of the metric and that taking these into account ``might alter the picture''.  Combining these two ideas, Haggard and Rovelli pointed out in \cite{haggard_quantum-gravity_2015} that when enough time has elapsed, quantum fluctuations of the metric might spark a `geometry transition' of a trapped region to an anti--trapped region. Then, the matter trapped inside the hole can escape. Bouncing black holes scenarios have been extensively considered elsewhere, in the context of resolving the central singularity and vis \`a vis the information loss paradox, see \cite{Malafarina:2017csn} for a recent review and also our concluding discussion.

The key technical result in \cite{haggard_quantum-gravity_2015} is the discovery of an `exterior metric' describing this process which solves Einstein's field equations exactly everywhere, except for the transition region which is bounded by a compact boundary. The existence of this exterior metric, which we henceforth refer to as the Haggard--Rovelli (HR) metric,\footnote{The HR metric is similar to the spacetimes considered in  \cite{barcelo_mutiny_2014,barcelo_lifetime_2015, barcelo_black_2016, barcelo_exponential_2016,hajicek_singularity_2001,
ambrus_quantum_2005-1}.} renders this process plausible: General Relativity need only be violated in a compact spacetime region, and this is something that quantum theory allows. The stability of the exterior spacetime after the quantum transition was studied in \cite{de_lorenzo_improved_2016}. The known instabilities of white hole spacetimes were shown to possibly limit the duration of the anti--trapped phase, but do not seem to otherwise forbid the transition from taking place. 

\begin{figure} 
\includegraphics[scale=0.4]{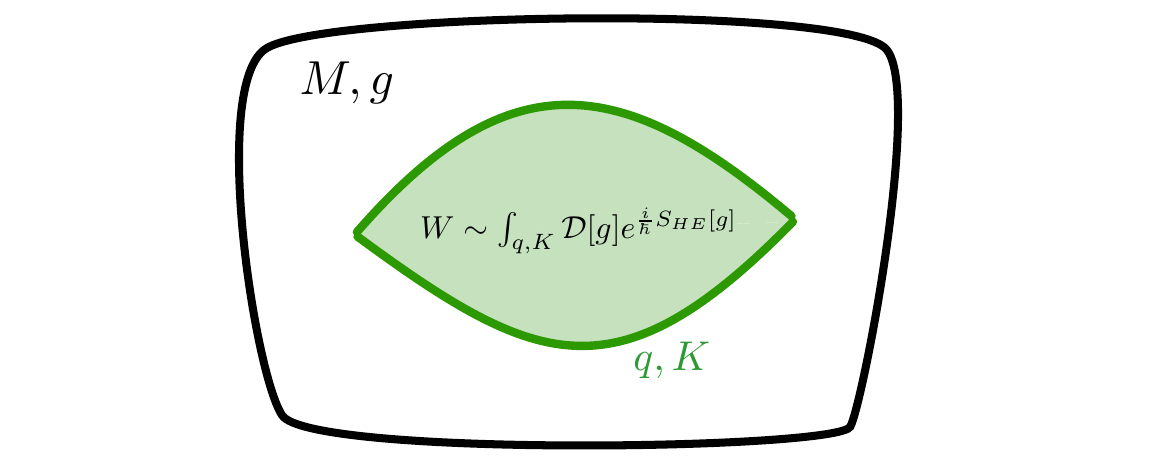} 
\caption{Geometry transition as a path integral over geometries. The shaded region (pale green) is where the quantum transition occurs. Outside this compact spacetime region, quantum theory can be disregarded and the geometry is a solution of Einstein's equations. This induces an intrinsic metric $q$ and extrinsic curvature $K$ of the boundary surfaces (dark green). The boundary state for the sum over geometries is a semiclassical state, peaked on both $q$ and $K$. The amplitudes of covariant LQG employed here display an emergent behavior as a Wheeler--Misner--Hawking sum in the limit of large quantum numbers.}
\label{fig:lens}
\end{figure}

The physics of the transition region can then be treated \`a la Feynman, in the spirit of a Wheeler--Misner--Hawking sum--over--geometries  \cite{misner_feynman_1957}, as sketched in Figure \ref{fig:lens}.  A theory for quantum gravity should be able to predict the probability of this phenomenon to occur and its characteristic time scales.

A first attempt to implement this program concretely was given in \cite{christodoulou_planck_2016} using the Lorentzian EPRL amplitudes in the context of covariant Loop Quantum Gravity (LQG). In this work, we complete the calculation laid out in \cite{christodoulou_planck_2016}. We estimate that the crossing time scales linearly with the mass. We show that in our setting and approximation the lifetime depends on the spread of the quantum state. The choice of a balanced semiclassical state gives an exponential scaling of the lifetime in the mass squared.  The calculation laid out in \cite{christodoulou_planck_2016} was built on a 2-complex without bulk faces. Our analysis is limited to this case, of 2-complexes without bulk faces. This is a significant limitation and in particular it is not clear how our estimates will be affected under refinements. The calculation presented here should be understood as a step towards a fuller calculation which should involve large 2-complexes with interior faces and ideally also a refinement procedure to explore the continuum limit.

Some improvements with respect to previous studies are the following. The assumption of a time symmetric process taken in \cite{christodoulou_planck_2016, haggard_quantum-gravity_2015} is dropped, allowing also for asymmetric processes as considered in \cite{de_lorenzo_improved_2016}. We do not choose an interior boundary surface that isolates the quantum transition (in \cite{christodoulou_planck_2016} a specific choice was made). The estimates are arrived at for an arbitrary interior boundary. We do not fix a specific 2--complex on which the spinfoam is defined. However, the technique we use is limited to a narrow class of spinfoam transition amplitudes, those defined on 2--complexes that do not have interior faces (which includes the amplitude considered in \cite{christodoulou_planck_2016}). As already discussed above, this is a significant limitation of the calculation and technique presented here. 

The paper is organized as follows. Before discussing the black hole case, in Section \ref{sec:tunneling} we review the case of a particle tunneling through a potential wall in non relativistic quantum mechanics and discuss the timescales involved in this process. In Section \ref{sec:RH} we review and improve the setup for the exterior spacetime, which describes the part of the spacetime well approximated by classical general relativity. We construct the exterior spacetime in a way that allows for the treatment of the phenomenon in a clear conceptual and technical setting. As explained in Section \ref{sec:RH} (see also Appendix \ref{sec:scalings}) the new construction of the exterior spacetime given here is adapted to the needs of the calculation and allows for a more clear conceptual setup. The relation to older constructions is explained in Appendices \ref{sec:crossedFingers}, \ref{sec:moreT},
\ref{sec:KruskalRH}.

 In Section  \ref{sec:W} we review the construction of the transition amplitude from covariant LQG and state the truncation/approximation used. In Section  \ref{sec:Bulk} we compute the characteristic time-scales using Loop Quantum Gravity for the class of 2-complexes with no interior faces and for an arbitrary choice of boundary surfaces. These results are confirmed numerically in Appendix \ref{sec:numerics} for the specific choice of boundary surface and 2-complex made in \cite{christodoulou_planck_2016}. The spinfoam calculation done here is based on the results of \cite{gravTunn}. In Section \ref{sec:shortcomings} we discuss limitations and shortcomings of our calculation. We close with a discussion of our results and comparison with other estimates appearing in the literature. 

\section{Tunneling timescales} \label{sec:tunneling}
We begin by reviewing the relevant timescales for a particle that tunnels through a potential wall in quantum mechanics. Consider a particle with energy $E$ that moves towards a potential barrier whose height is $V>E$.  Quantum theory predicts a probability $p$ for the particle to `tunnel through' the potential barrier. A good approximation to $p$ is given by 
\be
  p \sim e^{-\frac{\vert S_E \vert }{\hbar}},
\ee
which can be arrived at, for instance, using a saddle point approximation for the analytically continued path integral expression for the particle's propagator. Here, $S_E$ is the Euclidean action, which is in general complex.  There is no real solution of the classical equation of motion that crosses the barrier, but there is one after analytical continuation to the complex plane. Formally, this amounts to allowing the particle's velocity to become imaginary. The tunneling suppression exponent corresponds to the imaginary part of the action $S$, evaluated on the complex solution, and we define $S_E = i\, S$.

Suppose now that the potential barrier is a square barrier with height $V$, located in the region $0<x<L$. Imagine sending a wave packet that at time $T<0$ has a velocity $v>0$ (with mean kinetic energy $E<V$) and is centered at the position $x=v \,T <0$.  Around $T=0$ the packet hits the barrier and splits into a reflected packet with  an amplitude of modulus squared $1-p$ and a transmitted packet with an amplitude of modulus squared $p$. Suppose there is a detector on the other side of the barrier. The probability of this detector to detect the particle is $p$. But, what is the most probable time $T_c$ for the detector to detect the particle? The answer to this question defines the \emph{crossing time} for a tunneling phenomenon.  This is the time the actual tunneling takes to happen. Intuitively, it roughly corresponds to the classical flight time if the barrier was not present.   

Next, tunneling is the phenomenon that allows natural nuclear radioactivity. The radioactive decay of a nucleus can be modeled as a quantum particle trapped inside a potential barrier.  Imagine we have a wave packet with mean velocity $v$ bouncing back and forth inside a box of size $L$, whose walls are potential barriers of finite hight.  The particle will bounce against the wall with a period $\Delta T=L/v$. Thus, $\Delta T$ is a characteristic classical time of the phenomenon and at each bounce the wave packet has a probability $p$ to tunnel.  This implies that the probability to exit the barrier per unit time is $P \sim p/\Delta T$. The probability ${P}(T)$ for the particle to exit at time $T$ is then determined by $d{P}(T)/dT=-p \, P(T)$, namely 
\be
    {P}(t)=\frac1{\tau}\; e^{-\frac{t}{\tau}},
\ee
where 
\be
\tau \sim \frac{1}{P} \sim \frac{\Delta T}{p}\label{tau}
\ee
is the \emph{lifetime} of the nucleus. 

\bigskip

We have reviewed these simple physics to point out that we expect three distinct time scales at play.
\medskip
\begin{addmargin}[1em]{2em}
\emph{Lifetime $\tau$}: the time it takes a trapped particle to escape a trapping potential barrier. \\[1em]
\emph{Crossing time $T_c$:} the time needed to cross the potential barrier. \\[1em]
\emph{Characteristic time $\Delta T$:} the time that multiplies the inverse of the tunneling probability to give the lifetime. 
\end{addmargin}
\medskip
The crossing time $T_c$ and the lifetime $\tau$ are determined by quantum theory. They can be estimated from the propagator of the particle, contracted with coherent states $\ket{x,v}$ and $\ket{y,v}$ that are peaked on positions $x$ and $y$ left and right of the potential, respectively, and on a momentum given by a constant velocity $v$ and the mass of the particle:
\be
W(x,y,v;T)=\langle x,v|e^{-iHT/\hbar}|y,v \rangle,
\ee
where $H$ is the Hamiltonian. The crossing time can be estimated as the expectation value 
\be
T_c \sim \frac{\int_0^\infty dT\ T\ |W(0,L,v;T)|^2}{\int_0^\infty dT\  |W(0,L,v;T)|^2},\label{Tc}
\ee
which determines the average time after which the detector will click, \emph{when the tunneling takes place}.  The \emph{probability} of the tunneling to take place can be estimated from the amplitude of the propagator at this time
\be
p \sim |W(0,L;T_c)|^2,\label{Pt}
\ee 
and the lifetime $\tau$ follows from \eqref{tau}. The characteristic time $\Delta T$ is determined by the classical physical scales of the system, and is independent from $\hbar$. Further below, in Section \ref{sec:Lifetime}, we propose counterparts of these three time scales for the black to white  geometry transition.

\section{Haggard--Rovelli Spacetime} \label{sec:RH}
In this section we construct what we call here the Haggard-Rovelli spacetime. We follow a novel route for its construction that is adapted to the needs of the calculation and is more precise and conceptually clear. Note that the use of the word `spacetime' here is an abuse of terminology as this spacetime has a region missing, which is to be imagined as the slot where the LQG transition amplitude will go, see Figure \ref{fig:lens}. The important point will be that the exterior to this excised region geometry will be parametrised by two parameters, the bounce time $T$ and the mass $m$. There are four main regions in the spacetime, described by corresponding coordinate patches given explicitly below. With reference to Figure \ref{fig:ansatz}, region $I$ is the flat interior of a collapsing spherical shell. Region $II$ contains the trapped surface formed by the collapsing shell. Region $III$ contains the antitrapped surface formed by an expanding spherical null shell, while region $IV$ is the flat interior of this shell. 

Below we construct this spacetime step by step. We give a precise definition of the bounce time parameter and write the HR metric in a simple form that makes clear it should be understood as describing a two parameter family of exterior spacetimes. Note that the surface $\mathcal{Z}$ on which the junction condition is imposed between regions $II$ and $III$ is left arbitrary. 

The precise relation of the bounce time parameter with the extra parameters used in \cite{de_lorenzo_improved_2016,
bianchi_entanglement_2014}, is explained in Appendix \ref{sec:moreT}. These extra parameters encode useful geometrical properties of a given choice of interior boundary relating to the duration of the black and white hole phase, but do not affect the local bulk geometry and are not relevant for our calculation. We show how the presence of the horizon can be encoded in the boundary state through a scaling property of boost angles in the spacetime parameters in Appendix \ref{sec:scalings}. 

The logic followed to construct the exterior spacetimes is the following: we postulate the Penrose diagram on general considerations, and then show that the metric corresponding to this diagram indeed exists. The relation with the derivation in \cite{haggard_quantum-gravity_2015} is explained in Appendix \ref{sec:crossedFingers}.

\subsection{Global Structure}
\label{sec:globalHR}

The HR spacetime \cite{haggard_quantum-gravity_2015,de_lorenzo_improved_2016} constructed below provides a minimalistic model for a geometry where there is a transition of a trapped region (formed by collapsing matter) to an anti--trapped region (from which matter is released). The transition is assumed to happen through quantum gravitational effects that are non negligible only in a finite spatiotemporal region.

The transition region is excised from spacetime, by introducing a \textit{spacelike compact interior boundary}, which surrounds the quantum region. Outside this region the metric solves Einstein's field equations exactly everywhere, including on the interior boundary. 

\medskip

The HR spacetime is constructed by taking the following simplifying assumptions:
\begin{itemize}
\item Collapse and expansion of matter are modeled by thin shells of null dust of constant mass $m$. 
\item Spacetime is spherically symmetric. 
\end{itemize}

These assumptions determine the local form of the metric by virtue of Birkhoff's theorem, which can be stated as follows \cite{HawkEllis}: Any solution to Einstein's equations in a region that is spherically symmetric and empty of matter is \emph{locally} isomorphic to the Kruskal metric in that region. The HR spacetime is locally \emph{but not globally} isomorphic to portions of the Kruskal spacetime. 

Then, the metric inside the null shells is flat (Schwarzschild with $m=0$), the metric outside the shells is locally Kruskal with $m$ being the mass of the shells and spacetime is asymptotically flat. The trapped and anti--trapped regions are portions of the black and white hole regions of the Kruskal manifold, respectively. In particular, the marginally trapped and anti--trapped surfaces bounding these regions are portions of the $r=2m$ Kruskal hypersurfaces.

\begin{figure} 
\centering
\includegraphics[scale=.90]{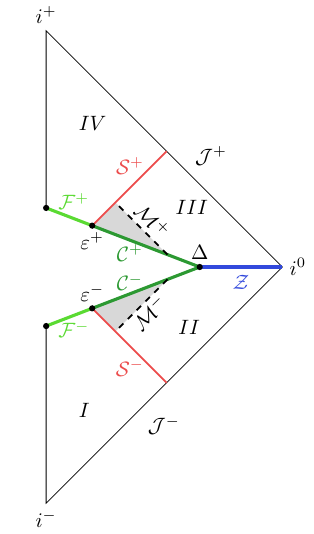} 
\caption{The Haggard--Rovelli family of `exterior spacetimes'. The collapsing null shell $S^-$ emerges as an anti-collapsing null shell $S^+$ after a quantum geometry transition. The shaded regions are (anti--) trapped. See Section \ref{sec:globalHR} for a detailed description. As explained in Section \ref{sec:BounceTime} this is is a two parameter family of exterior metrics.}
\label{fig:ansatz}
\end{figure}

It follows that the Carter--Penrose diagram of an HR spacetime is as shown in Figure \ref{fig:ansatz}. In particular, the HR metric must be such that the surfaces and regions in Figure \ref{fig:ansatz} have the following properties:

\begin{itemize}
\item $\mathcal{S}^-$ and $\mathcal{S}^+$ are null hypersurfaces. The junction condition on the intrinsic metric holds. Their interpretation as thin shells of null dust of mass $m$ follows: The allowed discontinuity in their extrinsic curvature results in a distributional contribution in $T_{\mu\nu}$ on $\mathcal{S}^-$ and $\mathcal{S}^+$, see next section. This is standard procedure in Vaidya null shell collapse models \cite{vaidya_gravitational_1951}, see for instance \cite{poisson2004relativist}. $T_{\mu\nu}$ vanishes everywhere else in the spacetime. 

\item The surfaces $\mathcal{F^+}$, $\mathcal{F^-}$, $\mathcal{C^+}$, $\mathcal{C^-}$ depicted in Figure \ref{fig:ansatz} are spacelike. Their union $\mathcal{B} \equiv \mathcal{F}^-\cup\mathcal{C}^-\cup\mathcal{C}^+\cup\mathcal{F}^+$ constitutes the interior boundary $\mathcal{B}$. The intrinsic metric is matched on the spheres $\Delta$ and $\varepsilon^{\pm}$. The extrinsic curvature is discontinuous on $\varepsilon^{\pm}$, see previous point, and is also discontinuous on $\Delta$ because of the requirement that $\mathcal{C^+}$ are spacelike: the normal to the surface jumps from being future oriented to being past oriented.

\item $\mathcal{Z}$ is a spacelike surface. The junction conditions for both the intrinsic metric and extrinsic curvature, hold, including on the sphere $\Delta$. As we will see below, $\mathcal{Z}$ plays only an auxiliary role and need not be further specified. See also Appendix \ref{sec:crossedFingers} for this point. 
\item $\mathcal{M^-}$ and $\mathcal{M^+}$ are marginally trapped (anti--trapped) surfaces and the shaded regions are trapped (anti--trapped). That is, the expansion of outgoing (ingoing) null geodesics vanishes on $\mathcal{M^-}$ ($\mathcal{M^+}$), is negative inside the shaded regions and positive everywhere else in the spacetime. 
\end{itemize}

Before explicitly giving the metric, let us comment on the necessity of extending the interior boundary outside the (anti--)trapped regions. By the assumption of spherical symmetry and Birkhoff's theorem, the marginally trapped and anti--trapped surfaces $\mathcal{M^-}$ and $\mathcal{M^+}$ can only be realized as being portions of the $r=2m$ surfaces of the Kruskal spacetime. If these do not meet the interior boundary, they must run all the way to null infinity. Thus, in order to have the global spacetime structure of a single asymptotic region, we must allow for non negligible quantum gravitational effects taking place in the vicinity, and crucially, outside, the (anti--)trapped surfaces. 

The metric, energy--momentum tensor and expansions of null geodesics are given in Eddington--Finkelstein coordinates in the following section. The metric is given in Kruskal coordinates in Appendix \ref{sec:KruskalRH}.

\subsection{ The exterior metric} \label{sec:EFRH}
In this section we explicitly construct the HR metric in Eddington--Finkelstein (EF) coordinates, in which it takes a particularly simple form. The union of the regions $I$ and $II$ of Figure \ref{fig:ansatz} is coordinatized by ingoing EF coordinates $(v, r)$ and the union of the regions $III$ and $IV$ by outgoing EF coordinates $(u, r)$. The junction on $\mathcal{Z}$ are given below. The radial coordinate $r$ will be trivially identified in the two coordinate systems. We work in geometrical units ($G=c=1$).

For the regions $I$ and $II$ the metric reads
\begin{eqnarray} \label{eq:downMetricEF}
\dd s^2 = -\left(1-\frac{2m}{r} \Theta(v-v_{\mathcal{S}^-}) \right)\dd v^2 + 2 \dd v \, \dd r+ r^2 \dd \Omega^2 , \nonumber \\
\end{eqnarray}
and for the regions $III$ and $IV$	
\begin{eqnarray} \label{eq:upMetricEF}
\dd s^2 = -\left(1-\frac{2m}{r} \Theta(u-u_{\mathcal{S}^+})\right)\dd u^2 -2\dd u\, \dd r + r^2 \dd \Omega^2 , \nonumber \\
\end{eqnarray}
where $\Theta$ is the Heaviside step function. The ingoing and outgoing EF times $v_{\mathcal{S}^-}$ and $u_{\mathcal{S}^+}$ denote the position of the null shells $\mathcal{S^-}$ and $\mathcal{S^+}$ in these coordinates. 

The two junction conditions on $\mathcal{Z}$ are satisfied by the identification of the radial coordinate along $\mathcal{Z}$ and the condition
\begin{equation} \label{eq:junctT}
v - u \stackrel{\mathcal{Z}}{=} 2r^\star(r) ,  \\
\end{equation}
where $r^\star(r)=r + 2m \log{\left|\frac{r}{2m}-1 \right|}$. Notice that this relation is the coordinate transformation between $(v,r)$ and $(u,r)$. We recall that the EF times are defined as $v=t+r^\star(r)$ and $u=t-r^\star(r)$, where $t$ is the Schwarzschild time.

 We emphasize that we need not and will not choose the hypersurface $\mathcal{Z}$ explicitly. The HR metric is independent of any such choice. The reason it is necessary to consider it formally as an auxiliary structure is that there does not exist a bijective mapping of the union of regions $II$ and $III$ of the HR spacetime to a portion of the Kruskal manifold. That is, it is necessary to use at least two \emph{separate} charts describing a Schwarzschild line element, as we did above. Where we take the separation of these charts to be (in other words, the choice of $\mathcal{Z}$), is irrelevant. See also Appendix \ref{sec:crossedFingers} for this point, in particular Figure \ref{fig:crossedFingers}.

A detail missing from previous works is that to explicitly define the metric we need to give the range of the coordinates for a given arbitrary choice of boundary surface.  This can be done as follows. Assume that an arbitrary but explicit choice of boundary surfaces $\mathcal{B}$ has been given and fixed. Having covered every region of the spacetime by a coordinate chart, we can describe embedded surfaces. Since all surfaces $\Sigma$ appearing in Figure \ref{fig:ansatz} are spherically symmetric, it suffices to represent the surfaces as curves in the $v-r$ and $u-r$ planes. Using a slight abuse of notation we write $v=\Sigma(r)$ or, in parametric form, $(\Sigma(r), r)$. The range of coordinates is given by the following conditions. For the regions $I$ and $II$ we have 
\begin{eqnarray} \label{eq:rangeDown}
&&v \in (-\infty, +\infty)\, ,\ \ \  r \in (0, +\infty)\, \nonumber \\ && 
 v \leq \mathcal{F^-}(r)\, ,\ \ \ v \leq \mathcal{C^-}(r)\, ,\ \ \  v \leq \mathcal{Z}(r)\, ,
\end{eqnarray} 
and for the regions  $III$ and $IV$ the coordinates satisfy
\begin{eqnarray}
&& u \in (-\infty, +\infty)\, ,\ \ \  r \in (0, +\infty)\, \nonumber \\ &&  u \geq \mathcal{F^+}(r)\, , \ \ \  u \geq \mathcal{C^+}(r)\, ,\ \ \  u \geq \mathcal{Z}(r).
\end{eqnarray} 

What remains is to ensure the presence of trapped and anti--trapped regions, as in the Carter--Penrose diagram of Figure \ref{fig:ansatz}. This is equivalent to the geometrical requirement that the spheres $\varepsilon^\pm$ have proper area less than $4 \pi (2m)^2$ while the sphere $\Delta$ has proper area larger than $4 \pi (2m)^2$. We may write this in terms of the radial coordinate as
\begin{eqnarray} \label{eq:radiusRequirements}
r_{\varepsilon^\pm} &<& 2m, \nonumber\\
r_{\Delta} &>& 2m.
\end{eqnarray} 
Apart from this requirement, the areas of the spheres $\varepsilon^\pm$ and $\Delta$ are left arbitrary. Since $\varepsilon^\pm$ and $\Delta$ are specified once the boundary is explicitly chosen, this is a condition on the allowed boundary surfaces that can be used as an interior boundary of a HR spacetime: $\mathcal{C^\pm}$ can be any spacelike surfaces that have their endpoints at a radius less and greater than $2m$, intersecting in the latter endpoint. Since $\mathcal{C}^\pm$ are spacelike, it follows that we necessarily have a portion of the (lightlike) $r=2m$ surfaces in the spacetime along with trapped and anti--trapped regions. See also Figure \ref{fig:fireworksPatch} for this point. The conditions
\begin{eqnarray} \label{eq:posDeltas}
v_\Delta & \geq & v_{\mathcal{S}^-}, \nonumber \\
u_\Delta  & \leq & u_{\mathcal{S}^+},
\end{eqnarray}
for the coordinates of the sphere $\Delta$ follow from equation \eqref{eq:radiusRequirements} and the fact that $\mathcal{C}^\pm$ are taken spacelike. 

\subsection{Exterior spacetime parameters}
The HR spacetime can be thought of as a two--parameter family of spacetimes with a compact portion `missing', in the following sense. The geometry of the spacetime, up to the choice of the interior boundary $\mathcal{B}$, is determined once two dimension--full, coordinate independent quantities are specified. One parameter is the mass $m$ of the null shells $\mathcal{S}^\pm$. The second parameter is the \emph{bounce time} $T$, the meaning of which is discussed in the following section. We can express $T$ in terms of $u_{\mathcal{S}^+}$ and $v_{\mathcal{S}^-}$ simply by 
\begin{equation} \label{eq:BounceTimeEF}
T = u_{\mathcal{S}^+}-v_{\mathcal{S}^-}.
\end{equation}
As with the mass $m$, the bounce time $T$ is taken to be positive. The positivity of $T$ is discussed in Appendix \ref{sec:moreT}.

Then, the Haggard--Rovelli geometry has \emph{two} characteristic physical scales: a length scale $G m/c^2 $ and a time scale $G T/c^3$, where we momentarily reinstated the gravitational constant $G$ and the speed of light $c$. The aim of this article is to compute the probabilistic correlation between the two scales $T$ and $m$ from quantum theory. This will be done in terms of a path integral in the region bounded by the interior boundary $\mathcal{B}$, with the boundary states peaked on the geometry of $\mathcal{B}$, without actually making an explicit choice for the hypersurfaces $\mathcal{C}^\pm$ and $\mathcal{F}^\pm$ that constitute the boundary $\mathcal{B}$.

\begin{figure} 
\centering
\includegraphics[scale=0.5]{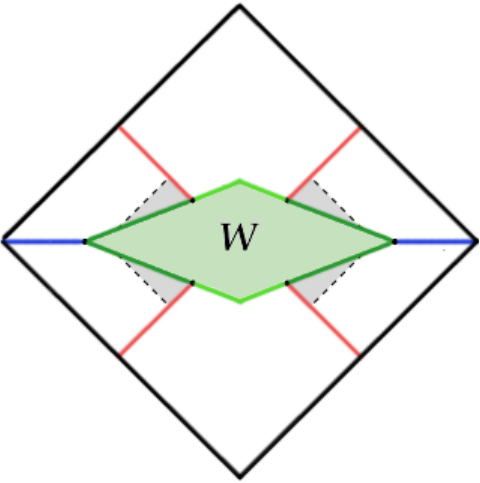} 
\caption{A cross--section of the rotated Carter--Penrose diagram of the HR spacetime, for easier comparison with Figure \ref{fig:lens}. The amplitude $W(m,T)$ gives the probability for the  spacetime with mass $m$ and bounce time $T$ to be realized.}
\label{fig:rotatedFireworks}
\end{figure}

The role of the bounce time $T$ as the second spacetime parameter is obscure in the line elements \eqref{eq:downMetricEF} and \eqref{eq:upMetricEF}. In equation \eqref{eq:BounceTimeEF}, we expressed the bounce time in terms of  the coordinate description of the collapsing and expanding (i.e.\! anti--collapsing) shells. The bounce time $T$ is then encoded implicitly in the line element via the Heaviside functions, which imply the inequalities $v \geq v_{\mathcal{S}^-} $ and $u \leq u_{\mathcal{S}^+}$ that specify the curved part of the spacetime.

We may make $T$ appear explicitly as a dimensionfull parameter in the metric components. This is achieved by shifting both coordinates $u$ and $v$ by 
\begin{eqnarray} \label{eq:timeShift}
v \rightarrow v- \frac{v_{\mathcal{S}^-}+u_{\mathcal{S}^+}}{2},\notag\\ 
u \rightarrow u- \frac{v_{\mathcal{S}^-}+u_{\mathcal{S}^+}}{2}.\end{eqnarray}
This is an isometry, since $(\partial_v)^\alpha$ and $(\partial_u)^\alpha$ are the timelike (piecewise, see next section) Killing fields in each region. It simply amounts to shifting simultaneously the origin of the two coordinates systems. The line elements \eqref{eq:upMetricEF} and \eqref{eq:BounceTimeEF} now read

\begin{eqnarray} \label{eq:downMetricNew}
\dd s^2 =  - \left(1- \frac{2m}{r}\,\Theta\left(v+\frac{T}{2}\right) \, \right)\dd v^2 + 2 \dd v \, \dd r+ r^2 \dd \Omega^2, \nonumber \\
\end{eqnarray}
and 
\begin{eqnarray} \label{eq:upMetricNew}
\dd s^2 = -\left(1-\frac{2 m}{r}\, \Theta\left(u-\frac{T}{2}\right) \, \right)\dd u^2 - 2 \dd v \, \dd r+ r^2 \dd \Omega^2. \nonumber \\
\end{eqnarray}

The role of $T$ as a spacetime parameter is manifest in the above form of the metric. It is instructive to compare it with the Vaidya metric for a null shell collapse model, describing the formation of an eternal black hole by a null shell $\mathcal{S}^-$ collapsing from past null infinity $\cal{J}^-$. Setting the shell to be at $v=v_{\mathcal{S}^-}$, the line element would be identical to \eqref{eq:downMetricEF}, with the difference that the range of the coordinates $(v,r)$ is not constrained by the presence of the surfaces $\mathcal{F^-}$, $\mathcal{C^-}$ and $\mathcal{Z}$, as in equation \eqref{eq:rangeDown}. The choice $v=v_{\mathcal{S}^-}$ for the position of the null shell is immaterial in this case and we can always remove $v_{\mathcal{S}^-}$ from the line element by shifting the origin as $v \rightarrow v-v_{\mathcal{S}^-}$. However, for the HR metric, the two coordinate charts are related by the junction condition \eqref{eq:junctT}. It is impossible to make both $v_{\mathcal{S}^-}$ and $u_{\mathcal{S}^+}$ disappear from the line element by shifting the origins of the coordinate charts, the best we can do is remove one of the two or, as we did above, a combination of them. This observation emphasizes that the bounce time $T$ is a free parameter of the spacetime. \footnote{The junction condition \eqref{eq:junctT} is unaffected by a simultaneous shifting of the form \eqref{eq:timeShift}.}

\subsection{Energy momentum tensor and expansions of null geodesics}

The energy momentum tensor and expansion of null geodesics has not been given in previous works. We report them here for completeness. It is straightforward to see that these are given by well known Vaidya expressions \cite{blau,poisson2004relativist}. For the energy momentum tensor we have 
\begin{center}
\medskip
\begin{tabular}{ r l }
  I $\cup$ II : &   $T_{\mu\nu} = + \frac{\delta(v+\frac{T}{2})}{4 \pi r^2} \delta^v_\mu \delta^v_\nu $, \\ & \\
  III $\cup$ IV : & $T_{\mu\nu} = - \frac{\delta(u-\frac{T}{2})}{4 \pi r^2} \delta^u_\mu \delta^u_\nu $.
  \end{tabular}
\medskip
\end{center}
The expansion $\theta^-$ of outgoing null geodesics in the patch $I$ $\cup$ $II$ and the expansion $\theta^+$ of ingoing null geodesics in the patch $III$ $\cup$ $IV$ read 
\begin{center}
\medskip
\begin{tabular}{ r l }
  I $\cup$ II : &   $\theta^- \equiv \nabla_\mu k_{-}^{\mu} = \Gamma^- \left(1- \frac{2 m}{r} \Theta(v+\frac{T}{2})\right) $,
   \\ & \\
  III $\cup$ IV : &   $\theta^+ \equiv \nabla_\mu k_{+}^{\mu} = -\Gamma^+ \left(1- \frac{2 m}{r} \Theta(u-\frac{T}{2})\right) $,
    \end{tabular}
\medskip
\end{center}
where $k_{-}^{\mu}$ and $k_{+}^{\mu}$ are affinely parametrized tangent vectors of the null geodesics and $\Gamma^\pm$ are the positive scalars making an arbitrary tangent vector $\tilde{k}^\alpha$ to a curve be affinely parametrized by defining $\tilde{k}^\alpha = \tilde{k}^\alpha$. From these expressions, it follows that the spacetime possesses a trapped and an anti--trapped surface, defined as the locus where the expansions $\theta^-$ and $\theta^+$ vanish respectively, and which we identify with $\mathcal{M^-}$ and $\mathcal{M^+}$ in Figure \ref{fig:ansatz}. Thus, in EF coordinates, $\mathcal{M^\pm}$ are given by 
\begin{center}
\medskip
\begin{tabular}{ l l l}
  $\mathcal{M^-}$ : &   $r=2m$ &, $v\in \left(-\frac{T}{2}, \,\mathcal{C^-}(2m)\right)$, 
   \\ & \\
  $\mathcal{M^+}$ : &   $r=2m$ &, $u\in \left(\mathcal{C^+}(2m), \,\frac{T}{2}\right)$.
    \end{tabular}
\medskip
\end{center}
As explained above, by the requirement $r_{\varepsilon^\pm} < 2m$ and $r_{\Delta} > 2m$, it will always be the case that the surfaces $\mathcal{M^\pm}$ are present in the spacetime, along with trapped and anti--trapped regions where $\theta^\pm$ are negative. We may explicitly describe the  trapped region as the intersection of the conditions $r<2m$, $v\in (-T/2,\mathcal{C^-}(2m))$, and $v \leq \mathcal{C^-}(r)$. Similarly, the anti--trapped region is given by  $r<2m$, $u\in \left(\mathcal{C^+}(2m),T/2\right)$ and $u \geq \mathcal{C^+}(r)$. The expansions $\theta^\pm$ are positive in the remaining spacetime.

\subsection{Bounce Time $T$} \label{sec:BounceTime}
The bounce time $T$ is a time scale that characterizes the geometry of the HR spacetime. Intuitively, $T$ controls the time separation between the two shells. In this section we discuss the meaning of $T$ as a spacetime parameter. We emphasize once again that $T$ is independent of the choice of interior boundary $\mathcal{B}$ and the choice of the junction hypersurface $\mathcal{Z}$ which are left arbitrary in our definition of the HR spacetime.

In equation \eqref{eq:BounceTimeEF}, we expressed the bounce time in terms of the null coordinates labelling the collapsing and expanding shells. As explained in \cite{haggard_quantum-gravity_2015}, the bounce time $T$ has a clear operational meaning in terms of the proper time along the worldline of a stationary observer (an observer at some constant radius $R$) who measures the proper time $\tau_R$ between the events at which the worldline intersects the collapsing and expanding shells $S^\pm$. A straightforward calculation yields 
\begin{equation} 
\tau_R= \sqrt{f(R)} \, \big(\, u_{\mathcal{S}^+}-v_{\mathcal{S}^-} + 2 r^\star(R) \, \big),
\end{equation}
where $f(R)=1-\frac{2m}{R}$. \footnote{Note that to get this expression we must add the contributions from the two line elements \eqref{eq:downMetricNew} and \eqref{eq:upMetricNew}, and use the junction condition  \eqref{eq:junctT}.} Using equation \eqref{eq:BounceTimeEF}, we have
\begin{equation} \label{eq:bouncetimeR}
T = \frac{\tau_R}{\sqrt{f(R)}}-2 r^\star(R). 
\end{equation}
Thus, the bounce time $T$ may be measured by an observer, provided they have knowledge of the mass $m$ and of their (coordinate) distance $R$ from the hole. 

The definition of $T$ as given in \cite{haggard_quantum-gravity_2015} was through the following expression
\begin{equation}  \label{eq:physMeanT}
T \approx \tau_R - 2R + \mathcal{O}\left( m \log \frac{R}{m}\right),
\end{equation}
which holds for for $R \gg m$.
 This approximate expression \eqref{eq:physMeanT} clarifies the physical meaning of $T$. As explained in \cite{haggard_quantum-gravity_2015}, for a far--away inertial observer and to the leading order in $R$, the bounce time $T$ corresponds to the `delay' in detecting the expanding null shell, compared to the time $2R$ it would take for it to bounce back if it were propagating in flat space and was reflected at $r=0$. To see this more clearly, we can introduce the dimensionless number $\tilde{R} \equiv R/m$ and  bring back $c$ and $G$. The bounce time $T$ can be measured through
\begin{equation} 
T \approx \tau_{R} - 2\tilde{R} \, \frac{G m}{c^2} + \mathcal{O}\left( \frac{G m}{ c^2} \log \tilde{R}\right), 
\end{equation}
which is a good approximation as long as $\tilde{R} \gg 1$.

Instead of the above approximate expression we will now use the exact formula \eqref{eq:bouncetimeR} which defines the bounce time $T$. Let us rephrase equation \eqref{eq:bouncetimeR} in order to see that $T$ is best understood as a spacetime parameter, a coordinate and observer independent quantity, and how it relates with the presence of symmetries of the spacetime. The exterior spacetime described by the HR metric has the three Killing fields of a static spherically symmetric spacetime, a timelike Killing field generating time translation and two spacelike Killing fields that together generate spheres. To be precise, these are piecewise Killing fields defined in each of the four regions of Figure \ref{fig:ansatz} and the spacetime is dynamical, not static, because of the presence of the distributional null shells $\mathcal{S}^\pm$. The orbits $\Upsilon$ of the timelike Killing field are labelled by an area $A_\Upsilon$: The proper area of a sphere generated by the two spacelike Killing fields on any point on $\Upsilon$. This is of course the geometrical meaning of the coordinate $r$. 

We can thus avoid to mention any coordinates or observers and specify $T$ through the following geometrical construction. Consider any orbit $\Upsilon$ that does not intersect with the interior boundary surfaces $\mathcal{B}$. The proper time $\tau_\Upsilon$ is an invariant integral evaluated on the portion of $\Upsilon$ that lies between its intersections with the null hypersurfaces $S^\pm$. For any such $\Upsilon$, we have

\begin{equation} \label{eq:defBounceTime}
T = \frac{\tau_\Upsilon}{\sqrt{f(A_\Upsilon)}}-r^\star(A_\Upsilon). 
\end{equation}   
Equation \eqref{eq:defBounceTime} should be read as follows: for any choice of $\Upsilon$ on the right hand side we calculate one quantity  $T[\Upsilon]$ on the left hand side. In principle, $T[\Upsilon]$ could depends on $\Upsilon$ since the right hand side depends on $\Upsilon$, but, it turns out that for all $\Upsilon$ we have that $T[\Upsilon]=T$. In this sense, the bounce time $T$ is independent of the chosen orbit $\Upsilon$. Also, note that it can be expressed only in terms of invariant quantities -- a proper area and a proper time. Expression \eqref{eq:defBounceTime} can be taken to be the \emph{definition} of $T$.

The bounce time $T$ can be understood in a couple more more ways which we note here for completeness and refer to the Appendices for further details. 
To link the above to the construction of the exterior metric in \cite{haggard_quantum-gravity_2015} we need only observe that the radius $r_\delta$ defined by $T=2r^\star(r_\delta)$ is where the null shells cross when the HR spacetime is mapped on the Kruskal manifold. This point is explained in detail in Appendix \ref{sec:crossedFingers}. The bounce time $T$ can also be understood as a time interval at null infinity, in analogy to an evaporation time, and can also be related to the parameters introduced in \cite{bianchi_entanglement_2014, de_lorenzo_improved_2016}, which admit a precise geometrical meaning as bounding the possible duration of the black and white hole phase as seen from infinity. This is explained in Appendix \ref{sec:moreT}

 We will discuss in Section \ref{sec:BoundaryState} that the scaling of boost angles with $m$ and $T$ encodes the presence of the (anti--) trapped surfaces $\mathcal{C}^\pm$  in the  semiclassical boundary state. In particular, the (anti--) trapped surfaces $\mathcal{C}^\pm$ can be equivalently characterized as the locus where boost angles do not scale with either $m$ or $T$. The scaling of geometrical invariants such as areas and angles in $m$ and $T$ is studied in Appendix \ref{sec:scalings}. Understanding the scaling of angles and areas with $m$ and $T$ is a main ingredient that allows to treat the geometry transition without specifying interior the boundary surface in the calculation of the crossing time and lifetime in Section \ref{sec:Bulk}.

\begin{center}
\noindent\rule{4cm}{0.4pt}
\end{center}

In summary, we have seen that the exterior spacetime described by the HR metric can provides a prototypical setup for studuing geometry transition. The geometry of the spacetime depends on two classical physical scales, which become encoded in the geometry of the interior boundary -- the boundary condition for the path integral. In turn, quantum theory is expected to correlate the two scales in a probabilistic manner. Let us now examine how this may be done using covariant Loop Quantum Gravity techniques.
 
\section{The transition amplitude $W(m,T)$} \label{sec:W}
Since the exterior geometry depends on the two parameters $m$ and $T$, so will the \emph{transition amplitude} $W(m,T)$ associated to the quantum region. This happens as follows. The exterior geometry induces an intrinsic geometry $q_{m,T}$ and an extrinsic geometry $K_{m,T}$ on the boundary $\mathcal{B}$. These depend on $m$ and $T$ since the full metric does. Let $\Psi_{m,T} \equiv \Psi[q_{m,T},K_{m,T}]$ be a coherent semiclassical state peaked on this 3d boundary geometry. Then,
\be \label{eq:theAmplitude}
W(m,T)= \langle W|\Psi_{m,T}\rangle
\ee
is the amplitude for the geometry transition where $\langle W \vert$ denotes the \emph{spinfoam amplitude map}. We invite the reader to compare Figure \ref{fig:rotatedFireworks} with Figures \ref{fig:lens} and \ref{fig:ansatz} for this point.

We recall that quantum gravity states cannot in general be split into an ``in'' and ``out'' state. This is the case here: The intrinsic and extrinsic geometry at the sphere $\Delta$ belongs to both surfaces $\mathcal{C}^-$ and $\mathcal{C}^+$. Since the state $\vert \Psi_{m,T}\rangle$ is peaked on the geometry of the entire boundary $\mathcal{B}$, it cannot be decomposed as $\vert \Psi_{m,T}\rangle \propto \vert \Psi^{\mathcal{C^-}}_{m,T}\rangle \otimes \vert \Psi^{\mathcal{C^+}}_{m,T}\rangle^\dagger $. The amplitude map acts instead on a generalized boundary state \cite{oeckl_general_2003,oeckl_general_2005}. Indeed, the crucial boost angle data for the transition are encoded on the sphere $\Delta$ which belongs neither to the future nor the past of the boundary surface $\mathcal{B}$.

 Formally, the transition amplitude can be written as a path integral over 4d geometries for a given boundary 3d geometry, conditioned on a semiclassical state peaked on both the intrinsic and extrinsic geometry of the boundary, see Figure \ref{fig:lens}.  Concretely, covariant Loop Quantum Gravity provides explicit formulas for spinfoam amplitude maps $\langle W|$ and for coherent states $\vert \Psi_{m,T}\rangle$ as we will see further below. Before that, let us discuss the relation between $W(m,T)$ and the timescales of the quantum transition.

\subsection{Timescales} \label{sec:Lifetime}

We consider a given black hole formed by collapse and estimate the characteristic time scales by quantum theory. That is, we \emph{fix the mass} $m$ and study how the quantum theory correlates the mass with the bounce time $T$, which is left arbitrary. Since the classical equations of motion are violated in the transition region we expect it is will be characterized by the timescales discussed in Section \ref{sec:tunneling}.   

We take the analog of the characteristic time of the phenomenon to be here simply the mass $\Delta T=m$ (in geometrical units, $G=c=1$). Since the mass $m$ is the only fixed physical scale in our problem and because $\Delta T$ is a classical quantity which cannot depend on $\hbar$, this is the only possible choice for the time scale $\Delta T$. It corresponds to the order of magnitude of the ``available time'' in the interior of the hole: We recall that the proper time along in--falling timelike trajectories, calculated from the (here, apparent) horizon to the singularity, is bounded from above by $\pi m$. We can imagine dividing the bounce time $T$ in intervals of order $m$ and ask what is the probability $p$ for the `tunneling' to occur in a single interval. This will give the lifetime $\tau$. Furthermore, we can ask what is the time the process itself takes, when it happens.  This is going to be the crossing time $T_c$. 

In Section \ref{sec:tunneling} we discuss how estimates for these $T_c$ and $\tau$ can be read from the transition amplitude \eqref{eq:theAmplitude}, a functional of the boundary geometry. The estimates for the characteristic time scales should be independent from the choice of interior boundary $\mathcal{B}$: the predictions of quantum theory are independent from where we set the boundary between the quantum and the classical systems, provided that the choice is such that the classical system does not include parts where quantum phenomena cannot be disregarded. 

Referring to the discussion of Section  \ref{sec:tunneling} we define the timescales relevant to the transition as follows. The crossing time is the mean value of $T$
\be \label{Tc2}
T_c \sim \frac{\int dT\ T\ |W(m,T)|^2}{\int dT  |W(m,T)|^2}.
\ee
The scaling of the tunneling probability $p$ with $m$ and $T$ can be estimated from the transition amplitude by setting $T=T_c$,
\be
p \sim |W(m,T_c)|^2.
\ee 
The lifetime is then defined as 
\be \label{eq:lifetimeFormula}
\tau \sim m \, |W(m,T_c)|^{-2}.
\ee 
These are the main formulas we use below to extract the relevant time scales from the transition amplitude $W(m,T)$.  
We emphasize that the above equations should be taken as the \emph{definition} of the quantities we seek to estimate here. While we have motivated the definition from an analogy with a quantum mechanical tunneling phenomenon, the analogy is far from perfect and it should be noted that the physical interpretation of quantum gravity amplitudes is not well established, contrary to quantum mechanical amplitudes. In particular, note that we have not normalised the probability considered above and it would not be clear how to do so. We are  assuming that the scaling of the probability with $T$ will be unaffected by the normalisation (similarly to a quantum mechanical tunneling ). We revisit this and other issues in Section \ref{sec:shortcomings}.

\subsection{Spinfoam Amplitude} \label{sec:spinfoams}

The spinfoam amplitude maps $\langle W \vert$ of covariant LQG \cite{rovelli_covariant_2014,perez_spin_2012,baez_spin_1998}, provide a tentative definition for the regularized path integral over histories of the quantum geometries predicted by LQG \cite{ashtekar_introduction_2013,thiemann_modern_2007,rovelli_quantum_2004,gambini_first_2011} to be the states of the quantum gravitational field.

 A spinfoam model is defined by a spin state--sum model, which defines the regularized partition function. The regularization is accomplished by a skeletonization on a 2--complex $\mathcal{C}$, a certain kind of topological 2--dimensional graph, with the sum over quantum geometries performed by a sum over spin configurations coloring the faces of $\mathcal{C}$ and its boundary graph $\Gamma$. 

 These quantum numbers label irreducible unitary representations of the Lorentz group, and recoupling invariants intertwining between them. They are interpreted as the degrees of freedom of the quantum gravitational field. The 2--complex $\mathcal{C}$ serves as a combinatorial book-keeping device, providing a notion of adjacency for a finite subset, a truncation, of the degrees of freedom of the quantum gravitational field.
 
 Starting with the Ponzano--Regge model \cite{ponzano_semiclassical_1969, regge_general_1961}, a progression through models defined in a variety of simplified settings \cite{ooguri_partition_1992,rovelli_basis_1993,turaev_state_1992,
boulatov_model_1992} culminated  within the framework of LQG to what has become known as the EPRL model \cite{barrett_lorentzian_2000,livine_new_2007,engle_flipped_2008,
freidel_new_2008,baratin_group_2012,dupuis_holomorphic_2012,engle_spin-foam_2013}, that treats the physically pertinent Lorentzian case. The EPRL amplitudes give a meaning to the formal expression 
\begin{equation}
 W = \int \mathcal{D}[\omega]\, \mathcal{D}[e]\; e^{i S_{H}[\omega,e]}.
\end{equation}
Here, $S_H$ is the Holst action for General Relativity, where the spin connection $\omega$ and tetrad field $e$ are the dynamical variables. 

 The spinfoam quantization program has seen significant advances over the past decades. The semiclassical limit of EPRL amplitudes defined on a fixed 2--complex and when all spins are taken uniformly large is well studied and closely related to discrete General Relativity
\cite{
bianchi_semiclassical_2009,barrett_asymptotic_2009,barrett_quantum_2010,
barrett_lorentzian_2010,magliaro_emergence_2011,magliaro_curvature_2011,han_asymptotics_2012,
han_asymptotics_2013,han_semiclassical_2013,
han_path_2013,engle_lorentzian_2016,
han_einstein_2017,Bahr:2017eyi,bahr_investigation_2016}. The semiclassical limit of the model reproduces the two--point function of quantum Regge calculus \cite{bianchi_lorentzian_2012, shirazi_hessian_2016,bianchi_lqg_2009,
alesci_complete_2008,alesci_complete_2007,
bianchi_graviton_2006}. 

The main feature that allows the study of the semiclassical limit is that the spinfoam amplitudes $\bra{W_\mathcal{C}}$ can be brought to the form \footnote{Throughout this work, we are using a simplified notation for the spinfoam amplitudes and boundary states to avoid technical details that may make it difficult for the not versed reader to follow the calculation that follows. Detailed definitions are given in \cite{gravTunn}, see also \cite{mariosPhd}.}
\begin{equation} \label{eq:spinfoamAmp}
\bra{W_\mathcal{C}}=W_\mathcal{C}(h_\ell) = \sum_{\{j_f\}} \nu(j_f) \int \! \dd g \, \dd z \;\prod_f e^{j_f \, F_f(g,z;h_\ell)}.
\end{equation}
 The variables $g$ are $SL(2,\mathbb{C})$ group elements living on the edges of $\mathcal{C}$ and the variables $z$ are spinors living on faces of $\mathcal{C}$ and are also associated to vertices of $\mathcal{C}$. The spins $j_f$ and functions $F_f(g,z;h_\ell)$ are associated to faces of $\mathcal{C}$. The function $F_f(g,z;h_\ell)$ refers to the face $f$ and will include a dependence on $SU(2)$ elements $h_\ell$ living on the boundary graph $\Gamma$ when the face $f$ touches the boundary.

The fact that EPRL amplitudes take the form of equation \eqref{eq:tempContAmp}, where the spins $j_f$ appear only in a polynomial summation measure $\nu(j)$ and linearly in the exponents, allows to use a stationary phase approximation when all spins $j_f$ are taken to be uniformly large. That is, when $j_f = \lambda \, \delta_f$, where $\lambda \gg 1$ and $\delta_f$ are of order unity. While this is a somewhat special configuration, it is particularly suitable for the physical problem considered here, where a uniform area scale $\lambda$ is dictated by the exterior metric (the mass $m$).

Below, we use the asymptotics of the Lorentzian EPRL model to estimate the two quantities defined in \eqref{Tc2} and \eqref{eq:lifetimeFormula}, the  lifetime $\tau$ and the crossing time $T_c$. A large uniform scale is provided by the mass $m$ and so we set $\lambda \sim m^2/\hbar$.  We are considering macroscopic black holes and emphasize that $\lambda$ is large but \emph{finite}. For instance, for a solar mass black hole $\lambda \sim 10^{39}$ and for a lunar mass black hole  $\lambda \sim 10^{31}$. The crossing time and lifetime are estimated to the leading order in $m$. That is, we will not be taking an actual limit. When we use the phrase `semiclassical limit' it should be understood colloquially. A significant limitation of the calculation that follows is that it concerns only complexes without interior faces. We discuss this and other limitations in Section \ref{sec:shortcomings}. 

\subsection{Truncation and Boundary Data} \label{sec:Truncation}
A choice of  2--complex $\mathcal{C}$ can be tought of as a truncation of the degrees of freedom of covariant LQG. It acquires an emergent interpretation in the semiclassical limit as being dual to a triangulation of spacetime. In this paper we restrict to spinfoam amplitudes defined on a fixed 2--complex topologically dual to a 4d triangulation of spacetime. That is, all faces have one link $\ell \in \Gamma$ in their boundary graph. 

Furthermore, we restrict to transition amplitudes defined fixed 2--complexes $\mathcal{C}$ with no internal faces. This is sufficient to complete the calculation laid out in \cite{christodoulou_planck_2016}, which this article follows up on. The behaviour of the amplitudes under refinements (or summation over 2-complexes depending on the point of view) \cite{dittrich_discrete_2012,
dittrich_coarse_2016,rovelli_quantum_2012} is not considered here.  This might be thought of as working in the so--called vertex expansion approximation, where the use of a single, fixed 2-complex defines a transition amplitude assumed to capture only some of the relevant degrees of freedom to the process. However, the validity of this approximation is doubtful and a point of contention \cite{Oriti:2006se} which we do not attempt to address in this article. The calculation presented here should be seen as a step towards a fuller calculation that involves a 2-complex with interior faces and in a setting in which it will be possible to understand the behaviour of the amplitude on large 2-complexes with interior faces and study its behaviour under refinements. 
 
The transition amplitude $W(m,T)$ is given by the EPRL amplitude $W_{\mathcal{C}}$, contracted with the boundary coherent states of equation \eqref{eq:ExtrinsicStates}. The boundary states are defined on the boundary graph $\Gamma \equiv \partial \mathcal{C}$. The continuous intrinsic and extrinsic geometry of $\mathcal{B}$ is approximated by a 3d triangulation, a piecewise--flat distributional 3d geometry, which is topologically dual to $\Gamma$. The metric information is discretized and encoded in the geometry of the boundary tetrahedra. The discretization is achieved by the assignment to each triangle, corresponding to a link $\ell$ in the dual picture, of the following discrete geometrical data. The area $A_\ell$ of the triangle, a boost angle $\zeta_\ell$ which determines a local embedding of the two tetrahedra that share the triangle $\ell$, and two normalized 3d vectors $\vec{k}_{s(\ell)}, \vec{k}_{t(\ell)}$ that encode the normal to the triangle as seen from each tetrahedron. 

These classical data completely specify the intrinsic and extrinsic geometry of a piece--wise flat 3d simplicial manifold, i.e.\! they determine an embedded spacelike tetrahedral triangulation. The notation $s(\ell)$ and $t(\ell)$ for the 3d vectors is standard and stands for ``source'' and ``target'', for the two nodes $\nn=s(\ell)$ and $\nn=t(\ell)$ sharing $\ell$. It refers to an arbitrary choice of an orientation for the links $\ell$ in $\Gamma$. The transition amplitude is independent of the choice of orientation and it does not enter the calculations that follow. Fixing an orientation for the links $\ell$, the boundary data $A_\ell,\zeta_\ell,\vec{k}_{s(\ell)}, \vec{k}_{t(\ell)}$ specify the boundary states of equation \eqref{eq:ExtrinsicStates}, the construction of which is discussed in the following section. To simplify notation, we denote the 3d normal data $\vec{k}_{s(\ell)}$ and $\vec{k}_{t(\ell)}$ collectively as $\vec{k}_{\ell\nn}$.

For what follows, it will be important to keep track of dimensions and in particular of $\hbar$. All quantities appearing in the definition of the boundary state $\ket{\Psi_{\Gamma}}$, given in equation \eqref{eq:ExtrinsicStates} below, are dimensionless, and the same is true for the spinfoam amplitude of equation \eqref{eq:spinfoamAmp}. We introduce the numbers $\omega_\ell \equiv A_\ell/\hbar $ which we call the \emph{area data}. The boost angles $\zeta_\ell$ are called the \emph{embedding data}. We will be mainly concerned with these two kinds of boundary data, which are gauge invariant. 

We recall that the starting point for the canonical quantization of General Relativity in LQG is to write GR in terms of the Ashtekar--Barbero (AB) variables, the AB connection $A$ and the densitized triads $E$. In these variables and at the kinematical level, the theory has the structure of a Yang--Mills theory with $SU(2)$ as symmetry group. The 3d normals $\vec{k}_{\ell\nn}$ are calculated in a given $SU(2)$ gauge, corresponding to a choice of local triad frame. The classical data $\omega_\ell,\zeta_\ell, \vec{k}_{\ell\nn}$ are called the \emph{boundary data} and will depend on the mass $m$ and the bounce time $T$. See  \cite{christodoulou_planck_2016} for a calculation of the boundary data for an explicit choice of boundary surfaces $\mathcal{B}$ and 2--complex $\mathcal{C}$.

The truncation has the effect that the transition amplitude is periodic in the embedding data $\zeta_\ell$ with a period $4 \pi /\gamma$  \cite{charles_ashtekar-barbero_2015}. That is, the transition amplitude is a function of the boundary data and satisfies
\be 
W_\mathcal{C}(\omega_\ell, \zeta_\ell, \vec{k}_{\ell\nn},t)= W_\mathcal{C}(\omega_\ell, \zeta_\ell + 4 \pi /\gamma, \vec{k}_{\ell\nn},t),
\ee
where the \emph{semiclassicality parameter} $t$ is introduced below. This truncation artefact can be read from  equation \eqref{eq:ExtrinsicStates}. It is a consequence of the discretization and the fact that the AB connection $A$ is an $SU(2)$ connection. The holonomy $h$ of $A$ is an element of $SU(2)$, a compact group, and fails to encode arbitrary boosts that in general take values in $[0,\infty)$. 

A simple example in which this effect can be seen is the following. Consider an intrinsically flat spacelike hypersurface equipped with Cartesian coordinates $x^1,x^2,x^3$, which is flatly embedded along $x^1$ and $x^2$. In these coordinates, the extrinsic curvature has only one non zero component which we call $K_{3}$ and the spin connection $\Gamma(E)$ vanishes. Consider the holonomy $h$ of the AB connection along a curve $\Upsilon$ given by constant $x^1,x^2$. We have
\begin{equation}
h = \mathcal{P} \,e^{\int_\Upsilon \Gamma(E) + \gamma K } = e^{i \gamma \frac{\sigma_3}{2} \zeta  },
\end{equation}
where $\zeta = \int_\Upsilon \dd x^3 \, K_{3}(x^3)$ corresponds to a smearing of the extrinsic curvature along $\Upsilon$ and can be used as embedding data. Then, $h$ is periodic in $\zeta$ with a period $4 \pi /\gamma$.

The contraction of the spinfoam amplitude with the boundary state peaks the $SU(2)$ elements $h_\ell$ in equation \eqref{eq:spinfoamAmp} on holonomies such as $h$. The consequence of the truncation is then that the transition amplitude is meaningful only for boundary states build with embedding data $\zeta_\ell$ that satisfy
\begin{equation} \label{eq:boostTruncation}
0\leq \zeta_\ell \leq  \frac{4\pi}{\gamma}. 
\end{equation}
We further discuss this limitation in Section \ref{sec:shortcomings}.

\subsection{Coherent Boundary State} 

\label{sec:BoundaryState}
The first step in building $W(m,T)$ is to construct a ``wavepacket of geometry'', a semiclassical state peaked on both the intrinsic and extrinsic geometry of a discretization of the boundary $\mathcal{B}$. The boundary states we consider in this paper are the gauge variant version of the coherent spin network states. We will first give their definition and then make an analogy with the usual Gaussian wavepackets from Quantum Mechanics to provide intuition.

The boundary states are defined as
\begin{eqnarray}\label{eq:ExtrinsicStates}
	\Psi_{\Gamma;\, \omega_\ell, \zeta_\ell, \vec{k}_{\ell\nn}}^t(h_\ell) &=& \sum_{\{j_\ell \}} \left( \prod_\ell d_{j_\ell} \e^{-\left(j_\ell - \omega_\ell \right)^2 t \,+\, i \gamma j_\ell \, \zeta_\ell } \right)\times \nonumber  \\ &&\ \ \ \ \ \times \, \psi_{\Gamma;\, \vec{k}_{\ell\nn} }(j_\ell; h_\ell),
\end{eqnarray}
where $h_\ell \in SU(2)$, $d_j=2j+1$ and $\gamma$ is the Immirzi parameter, the fundamental parameter of LQG, which is proportional to the smallest non zero quantum of area. The states $\Psi_{\Gamma;\, \omega_\ell, \zeta_\ell, \vec{k}_{\ell}}^t(h_\ell)$ are a Gaussian superposition of the coherent states $\psi_{\Gamma;\, \vec{k}_{\ell\nn} }(j_\ell; h_\ell)$. The latter are peaked on the intrinsic geometry of the triangulation of $\mathcal{B}$. They can be written explicitly in terms of Wigner D--matrices as
\begin{eqnarray} \label{eq:LSgroup}
\psi_{\Gamma,  \vec{k}_{\ell\nn}}(j_\ell; h_\ell) &=& \prod_\ell \sum_{m_s m_t} D^{j_\ell}_{m_s j_\ell}(k_{s(\ell)}^\dagger) \; D^{j_\ell}_{m_t j_\ell}(k_{t(\ell)}) \times
 \nonumber  \\ &&\ \ \ \ \ \times D^{j_\ell}_{m_s m_t}(h_\ell),
\end{eqnarray}
where the $SU(2)$ group elements $k$ are chosen appropriately so as to encode the corresponding 3d normals, see chapter $4$ of \cite{mariosPhd} for details. The \emph{semiclassicality parameter} $t$ controls the width of the Gaussians over the spins in \eqref{eq:ExtrinsicStates} and will play an important role in what follows.

The states $\Psi_{\Gamma;\, \omega_\ell, \zeta_\ell, \vec{k}_{\ell\nn}}^t(h_\ell)$ are semiclassical states in the truncated kinematical state space of LQG. The gauge invariant version of these states, where $SU(2)$ gauge invariance at each node of $\Gamma$ is imposed, was systematically introduced in \cite{bianchi_coherent_2010}. In that work, it was shown that they correspond to the large spin limit of Thiemann's $SL(2,\mathbb{C})$ heat kernel states \cite{thiemann_gauge_2001-3,
thiemann_gauge_2001,thiemann_gauge_2001-4}, in the twisted geometry parametrization \cite{freidel_twisted_2010,freidel_twistors_2010}. This parametrization corresponds to the boundary data considered here up to the twist angle $\alpha_\ell$, a further parameter which at the classical level allows for tetrahedral triangulations that are not properly glued along their faces. The twisted geometry parametrization labels points in the classical phase space of discrete general relativity in terms of data that are easy to interpret in terms of holonomies and fluxes (discrete versions of the AB variables). The heat kernel states in turn provide an overcomplete basis of coherent states for the corresponding truncated boundary Hilbert space of LQG, $\mathcal{H}_\Gamma=L^2\left[SU(2)^L / SU(2)^N\right]$, where $\Gamma$ is a graph with $N$ nodes and $L$ links. The quotient stands for the $SU(2)$ gauge invariance imposed at each node.  The gauge invariant version of the states $\psi_{\Gamma;\, \vec{k}_{\ell\nn} }(j_\ell, h_\ell)$ are known as the Livine--Speziale states \cite{livine_new_2007}. When boundary states are contracted with a spinfoam amplitude $W_\mathcal{C}$, the $SU(2)$ invariance at the nodes is automatically implemented and we need not consider the gauge invariant versions here. Further details on how to construct these states and how they are related see \cite{gravTunn,mariosPhd} and references therein.

 It can be instructive to compare the coherent spin network states defined in \eqref{eq:ExtrinsicStates} with the usual Gaussian wavepackets of Quantum Mechanics, which are peaked on a position $x_0$ and momentum $p_0$. In the position representation and up to normalization, we have
\begin{equation} \label{eq:qmWavepacket}
\Psi^t_{x_0,p_0}(x) \propto \int \dd p \; e^{-(p-p_0)^2 t + i p \, x_0} \; \psi(p,x),
\end{equation}
with $\psi(p,x) = e^{-ipx}$. 

In equation \eqref{eq:ExtrinsicStates} the $SU(2)$ group elements $h_\ell$ correspond to the (quantized) holonomies of the AB connection $A$ and play the role of the position variable $x$. The AB connection is the configuration variable of the AB variables. Its holonomy encodes the embedding of a canonical surface, along with information on the intrinsic curvature, because the AB connection is the sum of the Levi--Civita connection $\Gamma(E)$ and the extrinsic curvature $K$. The twisted geometry parametrization encodes $\Gamma(E)$ in the twist angle $\alpha_\ell$, which can be absorbed in an appropriate phase choice in the boundary states, see \cite{christodoulou_planck_2016}. Such a choice is assumed to have been made and the twist angle $\alpha_\ell$ is henceforth disregarded. The discrete version of the extrinsic curvature is encoded in the boundary state \eqref{eq:ExtrinsicStates} via the embedding data $\zeta_\ell$, which are analogous to $x_0$ in equation \eqref{eq:qmWavepacket}.

The fluxes are the discrete version of the conjugate variables $E$ of the AB variables. They encode the remaining geometrical information at the classical level and correspond to directed areas. The spins $j_\ell$ correspond to the area eigenvalues of the fluxes and play the role of the momentum variable $p$. The spins in \eqref{eq:ExtrinsicStates} are peaked on the area data $\omega_\ell$ which are analogous to $p_0$ in \eqref{eq:qmWavepacket}. 

The states $\psi_{\Gamma, \vec{k}_{\ell\nn}}(j_\ell, h_\ell)$ play the role of the plane wave $\psi(p,x) = e^{-ipx}$, understood as an eigenstate of the position operator, sharply peaked on the position $x$ (intrinsic geometry) and completely spread in the momentum $p$ (extrinsic geometry). Finally, the factors $d_j$ in \eqref{eq:ExtrinsicStates} are analogous to the integration measure $\dd p$ in \eqref{eq:qmWavepacket}.

Before closing this section we comment on how the boundary data encode the presence of the trapped and anti--trapped surfaces $\mathcal{C}^\pm$ in a discretization of the boundary $\mathcal{B}$. As shown in Appendix \ref{sec:scalings}, boost angles in the HR spacetime are in general functions of $X \equiv T/m$, and scale monotonically with $X$ (as well as with $T$ and $m$ separately). Whether they increase or decrease with $X$, is dictated by the \emph{sign} of the Schwarzschild lapse function $f(r)=1-2m/r$. In other words, an equivalent characterization of the (anti--)trapped surfaces $\mathcal{C}^\pm$ is to define them as the locus where $\frac{\dd \xi}{\dd X} =0$, where $\xi$ is any boost angle. Thus, the presence of the (anti--)trapped surfaces $\mathcal{C}^\pm$ will be encoded by the inverse scaling behavior of the embedding data $\zeta_\ell$, when corresponding to a discretization of the extrinsic curvature for parts of the boundary $\mathcal{B}$ with radius either smaller or larger than $r=2m$.

\section{Crossing time and Lifetime} \label{sec:Bulk}

\subsection{Transition Amplitude}
The transition amplitude is obtained by contracting the EPRL spinfoam amplitude \eqref{eq:spinfoamAmp} with a boundary state \eqref{eq:ExtrinsicStates}:
\be \label{eq:transAmpli}
W_\mathcal{C}(\omega_\ell, \zeta_\ell, \vec{k}_{\ell\nn},t)=\langle W_\mathcal{C}\vert \Psi_{\Gamma;\, \omega_\ell, \zeta_\ell, \vec{k}_{\ell\nn}}^t \rangle.
\ee
 The contraction is performed in the holonomy representation  by integrating over the boundary $SU(2)$ elements $h_\ell$. Following \cite{gravTunn}, the transition amplitude takes the form 
\begin{align} \label{eq:tempContAmp}
W_\mathcal{C}(\omega_\ell,\zeta_\ell,k_{\ell\nn}, t) &= \sum_{\{j_\ell \}} \left( \prod_\ell d_{j_\ell} \e^{-\left(j_\ell - \omega_\ell \right)^2 t \,+\, i \gamma j_\ell \, \zeta_\ell } \right) \times \nonumber \\ &\times \int\! \dd g\,\dd z\, \prod_{\ell} e^{j_\ell\; F_\ell(g,z;k_{\ell\nn})}.
\end{align}

The function 
\begin{equation} \label{eq:partialAmp}
I(j_\ell,k_{\ell\nn} ) = \prod_{\ell} e^{j_\ell\; F_\ell(g,z;k_{\ell\nn})}.
\end{equation}
is called the \emph{partial amplitude}. Because we restrict attention to 2--complexes  $\mathcal{C}$ without internal faces which are topologically dual to simplicial triangulations, each face $f$ has exactly one link $\ell$. We exploited this fact in trading the face subscripts $f$ in equation \eqref{eq:spinfoamAmp} for the corresponding links $\ell$. 

The spins $j_\ell$ are peaked on the area data $\omega_\ell$, corresponding to the triangle areas $A_\ell=\omega_\ell\, \hbar$ of a triangulation of $\mathcal{B}$. We consider a triangulation such that all discrete areas scale with $m^2$, the natural area scale of the spacetime. That is, 
\begin{equation}
A_\ell = m^2 \, \hbar \, \delta_\ell,
\end{equation}
with the \emph{spin data} $\delta_\ell$ being numbers of order unity. The spin data $\delta_\ell$ can have a dependence on $T/m$.\footnote{This is the case for the boundary data in \cite{christodoulou_planck_2016}, see equation \eqref{eq:bouDataExpl} and the discussion on this point in Section \ref{sec:shortcomings}}. Thus, the area data $\omega_\ell$ will  be of the form
\begin{align} \label{eq:DimLessArea}
	\omega_\ell(m,T) = \lambda \;\delta_\ell(X),
\end{align}
with $\delta_\ell(X)$ numbers of order unity for all values of $X$ allowed by equation \eqref{eq:boostTruncation}, and where we have defined 
\begin{equation}
\lambda \equiv \frac{m^2}{\hbar} \quad\text{and}\quad X \equiv \frac{T}{m}.
\end{equation}

We show in Appendix \ref{sec:scalings} that indeed all proper areas in the HR spacetime will be of the form $m^2 \, \delta(X)$ with $\delta(X)$ some function of $X$. This follows also on dimensional grounds. The areas $A_\ell$ are the result of a classical discretization and thus, $\hbar$ can only enter as an overall constant corresponding to the choice of units. Recall that we are working in geometrical units ($G=c=1$), where length, time and mass all have dimensions $\sqrt{\hbar}$. Similarly, since the embedding data $\zeta_\ell$ are boost angles, they will be functions only of $X$,
\begin{equation}
\zeta_\ell = \zeta_\ell(X),
\end{equation} 
and the same is true for the 3d normal data $\vec{k}_{\ell\nn}$.

The semiclassicality parameter $t$ controls the coherence properties of the states. As can be seen from \eqref{eq:ExtrinsicStates}, it must be a small and positive dimensionless number. Following \cite{thiemann_gauge_2001-4,bianchi_coherent_2010}, it corresponds to a dimensionless physical scale of the problem and is thus proportional to a positive power of $\hbar$. Since the only fixed physical scale available here is the mass $m$, we assume that
\begin{equation} \label{eq:semiPower}
t = \frac{\hbar^{n/2}} {m^n} \; , \quad n > 0.
\end{equation}
The allowed values of $n$ from the requirement that the states are peaked on both conjugate variables are given below. Note that in principle one may wish to allow in the parameter $t$ a parameter $T$. We revisit this point in Section \ref{sec:shortcomings}.

\subsection{Asymptotic formulas}
Below, we estimate the crossing time $T_c$ and lifetime $\tau$ using the results of \cite{gravTunn}. We briefly recall the setup and main results of that work. The area data $\omega_\ell$ and 3d normal data $\vec{k}_{\ell\nn}$ are assumed to be Regge--like \cite{barrett_asymptotic_2009}. This means that $\omega_\ell$ and $\vec{k}_{\ell\nn}$ specify a piecewise flat geometry for the 4d simplicial triangulation dual to the 2--complex $\mathcal{C}$. We emphasize that this assumption does not involve the embedding data $\zeta_\ell$. It implies that there exists a critical point for the partial amplitude of equation \eqref{eq:partialAmp}, which corresponds to a classical discrete \emph{intrinsic} geometry. The intrinsic geometry specified by $\omega_\ell$ and $\vec{k}_{\ell\nn}$ may be Lorentzian, 4d Euclidean or degenerate. The latter case corresponds to 4--simplices with vanishing four--volume.

 The main result in \cite{gravTunn} is that for a transition amplitude as in \eqref{eq:transAmpli} and for given spin data $\delta_\ell$, 3d normal data $k_{\ell\nn}$ and embedding data $\zeta_\ell$ that satisfy \eqref{eq:boostTruncation}, we have the estimate
 \begin{align}\label{eq:AmpEstimation}
	W_{\mathcal{C}} & \approx  \lambda^M \mu(\delta_\ell) \left[ \sum_{\{s(\text{v})\}}\prod_\ell \e^{-\frac{\Delta_\ell^2 }{4t} + i \gamma \Delta_\ell \delta_\ell}  \right] \times \nonumber \\
	&\times \left( 1 + \mathcal{O}(\lambda^{-1}) \; \right),
\end{align}
where we defined the \emph{embedding discrepancy}
\begin{equation} \label{eq:DeltaPhi}
\Delta_\ell = \gamma \zeta_\ell-\beta \phi_\ell(\delta_\ell) + \Pi_\ell.
\end{equation}
This estimate is the result of a stationary phase approximation in  $\lambda$, after suitable manipulations of \eqref{eq:tempContAmp}. To avoid confusion, we emphasize that the critical points discussed below are those of the partial amplitude $I(j_\ell,k_{\ell\nn} )$ of equation \eqref{eq:partialAmp}, not of the transition amplitude \eqref{eq:tempContAmp}.

The half--integer $M$ depends on the rank of the Hessian at the critical point, determined by $\delta_\ell$ and $\vec{k}_{\ell \nn}$, and on the combinatorics of the 2--complex $\mathcal{C}$. The function $\mu(\delta_\ell)$ includes the evaluation of the Hessian at the critical point. The parameters $\beta$ and $\Pi_\ell$ account for the different types of possible simplicial geometries, and whether we are at a link $\ell$ dual to a triangle at a corner of the boundary where the time orientation flips i.e.\! at the sphere $\Delta$ of Figure \ref{fig:ansatz}. This is called a thin--wedge. When $\delta_\ell$ and $\vec{k}_{\ell \nn}$ specify a Euclidean 4d geometry we have $\beta=1$ and $\Pi_\ell=0$. When they specify a Lorentzian geometry we have $\beta=\gamma$, $\Pi_\ell=\pi$ on thin--wedges and $\Pi_\ell=0$ otherwise. When they specify a 3d geometry we have $\beta=0$ and $\Pi_\ell$ is as in the Lorentzian case. As we will see below, the estimates for the scaling of the crossing time $T_c$ and the lifetime $\tau$ with the mass $m$ are independent of the above, and in particular do not depend on the type of discrete intrinsic geometry specified by $\delta_\ell$ and $\vec{k}_{\ell \nn}$. \footnote{The boundary data calculated in \cite{christodoulou_planck_2016} correspond to the degenerate type, see Appendix \ref{sec:numerics} for details.}

Each critical point of the partial amplitude comes with a $2^V$ degeneracy, corresponding to the different configurations for the orientation $s(v)$ of the tetrad, where $s(v)$ takes the value $+1$ or $-1$ on each vertex of $\mathcal{C}$. All $2^V$ critical points for given $\delta_\ell$ and $\vec{k}_{\ell \nn}$ correspond to the same intrinsic (Regge) geometry. The presence of multiple critical points corresponding to the same asymptotic geometry gives rise to the sum over the configurations of $s(v)$ in the estimate \eqref{eq:AmpEstimation}. This is a well known property of spinfoam models, see for instance \cite{rovelli_discrete_2012-1,christodoulou_how_2012, immirzi_causal_2016,vojinovic_cosine_2014}. It reflects the fact that the starting point for such models are tetradic actions such as the Palatini and Holst action for General Relativity, and not the Einstein--Hilbert action. The co--frame orientation $s(v)$ corresponds to the emergence of the discrete equivalent of the sign of the determinant of the tetrad field in the semiclassical limit. The Palatini deficit angle $\phi_\ell(\delta_\ell)$ depends also on $s(v)$ and corresponds to the usual Regge deficit angle when $s(v)$ is uniform. That is, when $s(v)=1$ for all vertices of the 2--complex $\mathcal{C}$ or $s(v)=-1$ for all vertices of $\mathcal{C}$.
 
\subsection{Crossing time and lifetime}
\label{sec:tcandtau}
 We are now ready to estimate the crossing time $T_c$ and lifetime $\tau$. The main observations we need from the equations \eqref{eq:AmpEstimation} and \eqref{eq:DeltaPhi} are the following. The transition amplitude depends on the bounce time $T$ only through $X$,
\begin{equation}
W_\mathcal{C}(\omega_\ell, \zeta_\ell, \vec{k}_{\ell\nn},t) = W_\mathcal{C}(m,X),
\end{equation}
while the mass $m$ appears explicitly through $\lambda$ and $t$. 

Next, the sum over the orientation configurations $s(\text{v})$ can be neglected for the following reason. The product over links in \eqref{eq:AmpEstimation} gives an overall exponent
\begin{equation}
o_{s(v)}=\sum_\ell -\frac{\Delta_\ell^2 }{4t} + i \gamma \Delta_\ell \delta_\ell,
\end{equation}
for each $s(v)$ configuration. This has  a positive real part and is in general different for each configuration of $s(v)$. 
Denoting $W_{\text{full}}$ the amplitude estimate in \eqref{eq:AmpEstimation}, and $W_{\mathcal{C}}$ the estimate when keeping only the critical point with $s'(v)$ such that $o_{s'(v)}$ is maximal, we have
\begin{equation}
	\frac{W_\text{full}}{W_{\mathcal{C}}} \sim 1 + \e^{-h(\delta_\ell,\zeta_\ell)/t},
\end{equation}
with $h(\delta_\ell,\zeta_\ell)$ a function with a positive real part. Thus, equation \eqref{eq:semiPower} implies that the full amplitude is well approximated by keeping only the contribution from the dominant co--frame configuration in \eqref{eq:AmpEstimation}, see also \cite{bianchi_semiclassical_2009}. 

\footnote{Note that, instead of the EPRL model, we may use the ``proper vertex'' model \cite{engle_spin-foam_2013,shirazi_hessian_2016,engle_proposed_2013}, where only a single co--frame orientation configuration survives in \eqref{eq:AmpEstimation}, corresponding to the Regge case for which $s(v)=1$ at every vertex. As we have seen above, the dominant co--frame orientation configuration in the EPRL model can correspond to \emph{any} configuration for $s(v)$. Hence, as expected, the two models will differ in their predictions for the quantum corrections to the lifetime $\tau$ and crossing time $T_c$ estimates.}

We have 
 \begin{equation} \label{eq:squaredAmpEstim}
	\vert W_{\mathcal{C}} \vert^2 \approx  \lambda^{2M} \mu(\delta_\ell)^2 \e^{-\frac{\sum_\ell \Delta_\ell^2 }{2t}} \left( 1 + \mathcal{O}(\lambda^{-1}) \; \right).
\end{equation}
The amplitude is suppressed exponentially as $\hbar \rightarrow 0$, matching the naive expectation for a `tunneling' phenomenon, unless all embedding discrepancies $\Delta_\ell$ vanish. This cannot be the case because it would indicate the existence of 
an exact classical solution of the (discretized) theory, connecting a black hole in the past to a white hole in the future. 

Plugging the above estimate into equation \eqref{Tc2}, we obtain the following expression for the crossing time
\begin{equation} \label{eq:tauX}
T_c = m \frac{\int \dd X \; X \,  \mu(X) \; e^{-\frac{1}{2t} \sum_\ell \Delta_\ell^2(X) }}
{\int \dd X \; \mu(X) \; e^{-\frac{1}{2t}\sum_\ell \Delta_\ell^2(X)}},
\end{equation}
where the upper limit of the integration range is defined by \eqref{eq:boostTruncation}. Hence, 
\begin{equation} 
T_c = m \, f(\gamma,t),
\end{equation}
with $f(\gamma,t)$ some function of the semiclassicality parameter $t$ and the Immirzi parameter $\gamma$. The precise form of $f(\gamma,t)$ will in general depend on the details of the discretization. However, inspection of equation \eqref{eq:tauX} reveals that when the function $\sum_\ell \Delta_\ell^2(X)$ has a minimum, for some $X=X_0$, the crossing time $T_c$ is independent of the discretization details to the leading order in $m$. We assume such a minimum to exist. The crossing time $T_c$ is then given by
\begin{equation} \label{eq:estimate}
T_c = m \, X_0(\gamma) \left( 1 + \mathcal{O}(t)\right).
\end{equation}
The above estimate follows from a direct application of the steepest descent approximation in $1/t$. Note that because in this approximation the crossing time observable does not depend on the spread of the extrinsic coherent states, the same result would be arrived at in the limit where the extrinsic coherent states become intrinsic.\footnote{We thank an anonymous referee for this remark.}

The dependence of the lifetime $\tau$ on $m$ can then be read out from $|W(m,T_c)|^2$, as in equation \eqref{eq:lifetimeFormula}. Setting $T=T_c=m\, X_0(\gamma)$ from the estimate \eqref{eq:estimate}, we have
\be
p \sim|W(m,T_c)|^2 \sim e^{-\frac{\Xi}{t(m)}},
\ee
where we neglected the polynomial scaling $\lambda^{2M}$ and defined $\Xi = \sum_\ell \Delta_\ell^2(X_0(\gamma))$ for brevity. As noted previously, the constant $\Xi$ cannot be zero. 
 
The lifetime $\tau$ then depends on the semiclassicality parameter $t$, determining the quantum spread of the boundary state. More precisely, it determines the relative balance of the quantum spread of the conjugate variables. A precise calculation for the allowed values of $n$ in \eqref{eq:semiPower} is given in \cite{bianchi_coherent_2010}. A quick way to summarise these results is as follows. From the definition of the boundary states, the spread in the areas $A_\ell$ and the embedding data $\zeta_\ell$ is
\be
 \Delta \zeta_\ell \sim \sqrt{t} \; ,\quad \Delta A_\ell \sim \frac{\hbar G}{\sqrt{t}},
\ee
where we have restored $G$ for clarity. In order for the state to be semiclassical we need both of these spreads to be small with respect to the corresponding expectation values. That is, $\Delta \zeta_\ell \ll 1$ and $\Delta A_\ell \ll A_\ell \sim m^2$. Thus,
\be
\frac{\hbar G}{m^2} \ll \sqrt{t} \ll 1.
\ee
Together with equation \eqref{eq:semiPower} this implies for $n$
\begin{equation} \label{eq:semiConstraints}
0 < n < 4.
\end{equation}
Taking the geometric mean for a balanced semiclassical state, this gives
\be
t = \frac{\hbar G}{m^2},
\ee
which in turn implies
\be
p \sim e^{-\frac{m^2}{\hbar G} \Xi}.
\ee
We have therefore recovered the naive semiclassical expectation for tunneling: the decay probability per unit of time $p$ is exponentially suppressed in a combination of the physical scales of the problem that has units of action. In the physical setup considered here, the only possibility would be a suppression in $m^2$. Finally, the resulting lifetime is 
\be\label{eq:LifeTimeResult}
\tau \sim m\  e^{\frac{m^2}{\hbar G} \, \Xi}.
\ee

 The scaling estimates for the crossing time $T_c$ and lifetime $\tau$ given in this section are analytic estimates for an arbitrary choice of boundary surface. They are verified numerically in Appendix \ref{sec:numerics}, for the explicit choice of hypersurfaces and discretization in \cite{christodoulou_planck_2016}. We now discuss the several limitations of the calculation. 

\section{Issues and Shortcomings}\label{sec:shortcomings} 
We now summarise and discuss several issues and shortcomings of the calculation. 

\emph{State Normalisation.} In writing the amplitude equation \eqref{eq:tempContAmp} we neglected a normalisation factor, see equations 32 and 33 of \cite{gravTunn}, because it depends only on $\omega_\ell$, it does not depend on the extrinsic data (boost angles) $\zeta_\ell$. The physics of the `geometry transition' studied here can be thought of as a change in the sign of the extrinsic curvature, which is encoded in the angles $\zeta_\ell$ corresponding to the sphere $\Delta$ (see also Appendix \ref{sec:scalings}).

However, that there can in general be a dependence on $T$ in $\omega_\ell$. An indication that this dependence may not be significant is given in Appendix \ref{sec:numerics}, where we consider the explicit choice of hypersurface made in \cite{christodoulou_planck_2016} (although we emphasize again that the calculation presented in this article considers an arbitrary boundary hypersurface $\mathcal{B}$). There is then indeed a dependence of $\omega_\ell$ on $T$ but this does not affect the result as the dependence of $\omega_\ell$ on $T$ is very weak. Using the values for the boundary data $\omega_\ell$ found in \ref{sec:numerics}, and since the amplitude is evaluated on $T=T_c$ and we find $T_c \sim m$, this just gives a factor of order unit in $\omega_\ell$. We also note that is not clear how one could encode in the area data the flip of the sign of the extrinsic curvature.

Having said the above, we note that it is possible that when considering large 2-complexes or a refinement limit the normalisation factor may be a subtle point to consider and may become relevant. 

\emph{Bounded range of values of boost angles.}
Our results hold only for boundary data satisfying equation \ref{eq:boostTruncation}. This implies that for fixed $m$ and $\gamma$, only spacetimes characterized
by certain values of $T$ are allowed. As explained in Section \ref{sec:BoundaryState},  this is a discretisation artefact related to the use of the Ashtekar-Barbero connection. It arises because of the encoding of a boost angle, an unbounded parameter of the Lorenz group, into an $SU(2)$ variable, which is a compact group. 

Naively, it would seem that if one were to consider refinements of the 2-complex (and therfore also of its boundary), larger total boost angles could be encoded because they could be split in many smaller boost angles. Then, a larger range of $T$ could be explored. On the other hand, there is also the possibility that in LQG there is a maximal boost angle allowed. For this, see for instance \cite{Rovelli:2013osa,charles_ashtekar-barbero_2015} and also Appendix E in \cite{MariosGeometryTransitionCovariant2018}. The behaviour of boost angles and in particular those at the sphere $\Delta$ under refinements is another subtle point that would need to be addressed in future works. 

\emph{Dependence of $t$ only on $m$.} We have assumed that the semiclassicality parameter $t$ depends only on the mass $m$. Because the logic followed here is to consider some fixed mass $m$ and allow $T$ to vary, this corresponds to considering semiclassical states with fixed spread. We are following \cite{bianchi_coherent_2010} on this, and the same was assumed in \cite{gravTunn} which contains the main spinfoam techniques used here. However, inn principle, one may consider a dependence on $t$ also on $T$ (or on $X \equiv T/m$).\footnote{We thank an anonymous referee for pointing this out.} It is not clear to us how the calculation may be affected if one allows for such a dependence of $t$ on $T$, nor how to determine what the appropriate dependence should be. More in general, because $t$ is crucial to estimate the lifetime, it would be important in future work to study this and hopefully arrive at a way to determine how it should depend on the spacetime parameters. 

\emph{Choice of balanced state.} 
Related to the above point, our calculation finds that the lifetime will depend on the spread of the quantum
state. On the other hand, its dependence on the mass can take a large range of values, see Section \ref{sec:tcandtau}. Other than choosing a balanced state in the area and boost angle variables, we do not have a better argument for what should be the chosen value of $t$. This is a clear indication that the approximations used here are not sufficient to estimate the lifetime with any certainty. This is another indication that it will be important in future work to understand better the role of $t$ in the calculation.

\emph{Independence on choice of boundary surface}
The boundary surface $\mathcal{B}$ may be thought of as a choice of a `Heisenberg cut'. In our calculation, we did not fix a choice of this interior boundary. The scaling estimates have been arrived at analytically based on general geometrical properties of the exterior spacetime. Indeed, it should be demanded that the predictions of the theory do not depend on such a fiducial choice.

This point was examined also in Appendix \ref{sec:numerics} and Appendix \ref{sec:scalings}. In Appendix \ref{sec:numerics} we confirmed that the result given in the main text (where we did not fix a specific hypersurface), is indeed reproduced when a specific choice of hypersurface is done. The specific hypersurface and discretisation considered in Appendix \ref{sec:numerics} is that of \cite{christodoulou_planck_2016}, which this work follows up on. In Appendix \ref{sec:scalings} we show that any boost angle between two timelike vectors will scale monotonically with $X \equiv T/m$, as well as with $T$ and $m$ separately. Since the transition concerns essentially the flipping of the sign of the extrinsic curvature encoded in the boost angles at the sphere $\Delta$, this gives a further argument that one can indeed hope to arrive at scaling estimates, such as the one we arrive at here, independently from the choice of hypersurface by considering geometrical properties of the exterior spacetime.

The independence of our estimates on the choice of $\mathcal{B}$ is an encouraging sign. But, we warn the reader this is a first result and not a generic conclusion, that should be read in the context of the specific task set out in this manuscript: to estimate the \emph{scaling} of the timescales we have defined with the mass.

In particular, while the scalings found here do not depend on the choice of hypersurface, in general it should be expected that numerical factors would depend on the boundary chosen, as well as on the choice discretisation of the boundary. Furthermore, if one were to consider a refinement limit, the behaviour of such factors may become of relevance. 

\emph{The analogy with tunneling.}
The analogy we have made in this manuscript with a tunneling phenomenon (see Section \ref{sec:tunneling} and Section \ref{sec:Lifetime}) should be understood as an intuitive inspiration for defining the relevant timescales we want to estimate in this prototype calculation that seeks to employ quantum gravity amplitudes to extract some physical observable. The correct interpretation of quantum gravity transition amplitudes such as the one we consider here is not well established (contrary to quantum mechanics), notwithstanding because any predictions that may be extracted from the theory have not been tested against experiment. 

We also note that the problem we study here seems reminiscent of the arrival time problem \cite{Aharonov:1997md,Delgado:1997tj,Grot:1996xu}.\footnote{We thank an anonymous referee for pointing us to relevant literature.} It is not clear to us whether a good analogy to this well studied problem and the problem we study here can be arrived at. As discussed in the introduction of Section \ref{sec:W}, here we do not have analogues of `in and an out' states. The main variable for the transition are the angles at the sphere $\Delta$ which belong neither to the `past' nor to the `future' of the transition. Also, it is not clear what would serve here as the `measurement' or `collapse' and indeed we would not here have an analogue of a `time' operator. Therefore, the analogy of the black to white hole transition with a quantum mechanical tunneling or the flight time problem is far from perfect. 

\emph{Transition to `other' geometries}
Another difficulty in interpreting the transition amplitudes we considered is that one may wonder whether the portion of spacetime with the trapped region may tunnel to `some other geometry' than an anti--trapped region according to the same theory, and ask how the `probability' of such an eventuality may compare to that of the black to white transition. We now briefly comment on this point although we caution that this work does not offer much insights on the possible answer to this question. 

 It can be noted that it is not obvious to us how to do such a comparison in the first place, at least with the techniques we used here. The calculation presented here relies on properties of the exterior family of metrics. We have considered the exterior spacetime as given in all spacetime except a compact region. It is not clear if one could define some timescale analogous to $T$ that would have a meaning for an arbitrary exterior spacetime, therefore rendering the entire calculation not applicable. Indeed, as we have seen in Section \ref{sec:BounceTime} here $T$ is best thought of as a parameter of the specific exterior geometry we consider. Therefore, it would seem one would need to attempt to compare `different geometry transitions' at the basis of their `absolute probabilities'. 
 
In other words, the calculation presented here is not how the probability of some geometry X transitioning to a geometry Y compares to the probability of geometry X transitioning to some geometry Z, (or, to no transition). We have not here defined a probability in the sense that one transition is normalised against the sum of the probabilities of all possible geometry transitions. Rather, the spirit of the calculation is best understood as estimating how the probability of transition \emph{scales} with the timescale $T$ 
and the mass scale $m$ having assumed this transition to take place. Then, an implicit assumption is that, like in quantum mechanical tunneling, one may hope to estimate how the probability scales with time without needing to calculate its normalisation.

In summary, the logic followed here is that the external geometry is fixed to be the HR spacetimes, not that there is an in state and various possible out states, or that there is the interior boundary $\mathcal{B}$ and one considers all possible spacetimes that could `continue' this boundary to asymptotic infinity. To do this, one would need to somehow be able to compare the probabilities between `one or the other geometry transition happening'. We do not know how this could be done. Certainly, in a full theory and a full treatment of the phenomenon, it should  be requested that such are  in principle possible. 

\emph{Interior faces and Refinements}
In this manuscript we have only considered amplitudes defined on 2-complexes without interior faces. The aim here was to complete the calculation laid out in \cite{christodoulou_planck_2016} where a very coarse 2-complex without interior faces was considered. The semiclassical limit techniques we used are explained in \cite{gravTunn}, which also only deals with 2-complexes without interior faces. 

To be clear, arbitrarily large 2-complexes with no internal faces can be constructed. But, these would not be  particularly interesting as they would in general correspond to peculiar configurations resembling long tubes. Imagine for instance stacking hypercubes along one direction to make a long rectangle: the dual graph will not have any internal faces. What is done here is applicable to such `tree level' complexes, where there is no summation over quantum numbers labelling an internal (`virtual') face. This is a very significant limitation of the calculation. Essentially, in this setting, the behaviour of the amplitude is largely determined by the boundary state (see also next paragraph). Indeed, a main challenge for future work would be to consider 2-complexes with internal faces.

\emph{Is there any `true' input from EPRL?}
Our calculation relies on the asymptotics of the EPRL model. This in the end comes down to using a Regge type on shell action. However, note that the asymptotics of the EPRL model give rise to an area--angle Regge action (not the original length--angle Regge action), and involve the Immirzi parameter in a non trivial way because of the Ashtekar-Barbero connection. For this, see equation 95 of \cite{gravTunn}, which is the main result of that work and which was used here to approximate the amplitudes (it corresponds to our equations \eqref{eq:AmpEstimation} and \eqref{eq:DeltaPhi}). On the boundary, the Immirzi parameter multiplies the boost angles, see also Section \ref{sec:BoundaryState}. Then, essentially the exponential supression of the transition amplitude is due to the mismatch of the boost angles as given by the asymptotics of the spinfoam amplitude and as encoded in the boundary state. 

Other than the above point, our calculation in a sense does not include what may be regarded as `true dynamical input' from the EPRL model specifically, as it is not sensitive to the bulk degrees of freedom of the model. Then, the estimates given here may be possible to be arrived at with a general line of argument on how a spin sum model based on the Ashtekar-Barbero variables should be expected to behave in the semiclassical limit. 

\section{Discussion and comparison with earlier results}\label{sec:Discuss} 

The calculation we presented completes the task set out in \cite{christodoulou_planck_2016} and is based on the techniques detailed in \cite{gravTunn}. We have defined and discussed the timescales characterizing the geometry transition of a trapped to an anti--trapped region and provided estimates using covariant Loop Quantum Gravity. The crossing time $T_c$ characterizes the duration of the process. We find that quantum theory suggests that it scales linearly with the mass. The lifetime $\tau$ is a much larger time scale, which we interpret as the time at which it becomes likely that the transition takes place. The geometry transition is governed by the boundary data on the `corner' of the lens region and may be understood as coming down to flipping the sign of the extrinsic curvature. One significant improvement from the calculation set out in \cite{christodoulou_planck_2016} is that we arrive at estimates of the crossing time and lifetime without fixing a specific boundary hypersurface. 

While the scaling of the crossing time $T_c \sim m$ appears to us now well established, the lifetime $\tau$ calculated here is only a preliminary result. Our results seem to favor an exponential scaling of the lifetime in the square of the mass $m$, in accordance with the na\"ive expectation for a tunneling phenomenon, but further investigations will be necessary before any conclusive result may be claimed. The many shortcomings of our calculation for the lifetime are discussed in the previous section.

We now give a brief comparison of relevant results in the existing literature.
A polynomial scaling for the lifetime $\tau$ in the mass $m$ was suggested in \cite{haggard_quantum-gravity_2015, christodoulou_planck_2016}, and phenomenological consequences were studied in   \cite{barrau_fast_2014,barrau_phenomenology_2016,
barrau_planck_2014,vidotto_quantum-gravity_2016}. To be clear, the possibility of a polynomial scaling has not been excluded here. In particular, this possibility is allowed by the bounds of equation \eqref{eq:semiConstraints}. 
 
Singularity resolution in black holes has been extensively studied in the canonical approach to LQG, see for instance \cite{modesto_black_2008,
gambini_introduction_2015,corichi_loop_2016} and references therein. Current investigations suggest singularity resolution through a bounce to a white hole, with characteristic time scales reported in \cite{corichi_loop_2016,olmedo_black_2017}. These 
studies are based on a canonical quantization of the trapped and anti--trapped regions and concern only the interior of the hole. In that line of work, the corresponding physics far from the transition region are less clear. On the contrary, when using the path integral approach, the details of the interior process are, strictly speaking, irrelevant. The two frameworks in this sense may be thought of as complimentary. Further developments are necessary before a comparison of the results from the covariant and canonical framework of LQG regarding the black to white hole transition seem possible.

Two lines of investigation outside the context of LQG have used an exterior spacetime closely related to the HR spacetime. The quantum transition of a trapped to an anti--trapped region has been studied by H\'aj\'i\v{c}ek and Kiefer in \cite{hajicek_singularity_2001}, using  an exact symmetry reduced null shell quantization scheme. The timing of the transition was subsequently studied by Ambrus and
H\'aj\'i\v{c}ek in \cite{ambrus_quantum_2005-1}. More recently, Barcel\'o, Carballo-Rubio and Garay studied the transition in a series of papers \cite{barcelo_mutiny_2014, barcelo_lifetime_2015, barcelo_black_2016, barcelo_exponential_2016}, by performing a Euclidean path integral in the quantum region.  Both these lines of investigation identify  a time scale that scales linearly with the mass $m$.  Our result for the crossing time $T_c$ corroborates these results. We have emphasized that the crossing time $T_c$ must not be confused with the lifetime $\tau$, see Section \ref{sec:tunneling}. The lifetime of the black hole is the expected time between the formation of the black hole and its quantum transition to a white hole. The crossing time is the (much shorter) time that characterizes the duration of the transition itself. 

There are two obvious reasons for which the lifetime $\tau$ cannot be of order $m$. The first is that the empirically established existence of black holes in the sky immediately falsifies any prediction for a lifetime $\tau \sim m$. The second reason is that a transition from a black hole to a white hole is forbidden in the classical theory, therefore the lifetime must go to infinity in the limit in which we take $\hbar$ to zero.  This is not the case if $\tau$ is proportional to $m$, because no $\hbar$ is present in this relation. This is clearly pointed out in \cite{ambrus_quantum_2005-1} by Ambrus and H\'aj\'i\v{c}ek, where the authors call their result $\tau\sim m$ ``unreasonable", and leave the question open. In our opinion, the distinction we have made between crossing time and lifetime  clarifies this issue. 

The technique we have used has several approximations, shortcomings and limitations. These are discussed in Section \ref{sec:shortcomings}. Regarding the calculation of the lifetime, which is the main quantity of interest, this work should be thought of as just one step and possibly a roadmap towards fuller calculations, rather than a prediction of covariant Loop Quantum Gravity.  

\section*{Acknowledgments}
The authors thank Carlo Rovelli, Tommaso de Lorenzo, Simone Speziale and Hal Haggard for the many valuable discussions and insights on this work, and Alejandro Perez and Abhay Ashtekar for several crucial critical remarks. The ideas leading to the results presented in this paper were influenced by insightful discussions with Beatrice Bonga, Abhay Ashtekar, Jorge Pullin and Parampreet Singh during a visit in Louisiana State University, Eugenio Bianchi during a visit at the Penn State University, Jonathan Engle and Muxin Han during a visit at the Florida Atlantic University, and Louis Garay and Raul Carballo--Rubio during a visit at the Complutense University of Madrid. We thank them for their input and for their hospitality. Jonathan Engle is also thanked for several subsequent communications and remarks.

 \section*{Chronology note}
 The work presented here was mainly done during the period 2016-2018. The technique on which the calculation of the estimate for the bounce time of the black to white transition presented here is based can be found in \cite{gravTunn} was based. Some of the details presented here can be found scattered in the PhD manuscripts of the authors, although different notation and conventions may be used. 

To avoid confusing the narrative, we did not cite above throughout the main body of the paper several works that have appeared after this work first appeared as a preprint. A list of such works for which the results presented here can be relevant is \cite{Gozzini:2021kbt, Dona:2022dxs, Dona:2022yyn, Dona:2023myv, 
Dona:2018nev,
Bianchi:2018mml, Vidotto:2018wvr, DAmbrosio:2018wgv, Rovelli:2018cbg, Rovelli:2018hbk, Rovelli:2018okm, Alesci:2018loi, Kiefer:2019csi, Martin-Dussaud:2019wqc, Schmitz:2019jct, BenAchour:2020bdt, Piechocki:2020bfo, BenAchour:2020gon, Ong:2020xwv, Kelly:2020lec, Kelly:2020uwj, Zhang:2020lwi, DAmbrosio:2020mut, Schmitz:2020vdr, Barrau:2021spy, Mele:2021hro, Munch:2021oqn, Ansel:2021pdk, Soltani:2021zmv, Rignon-Bret:2021jch, Husain:2022gwp, Barcelo:2022gii, Kazemian:2022ihc, Phat:2022xxw}.  

\medskip

MC acknowledges support from the SM Center for Spacetime and the Quantum and from the Educational Grants Scheme of the A.G. Leventis Foundation.

\appendix

\section{Lifetime and Crossing Time for the Boundary Data of the Setup in \cite{christodoulou_planck_2016}} \label{sec:numerics}
\begin{figure}
\begin{tabular}{lr}
\includegraphics[scale=0.155]{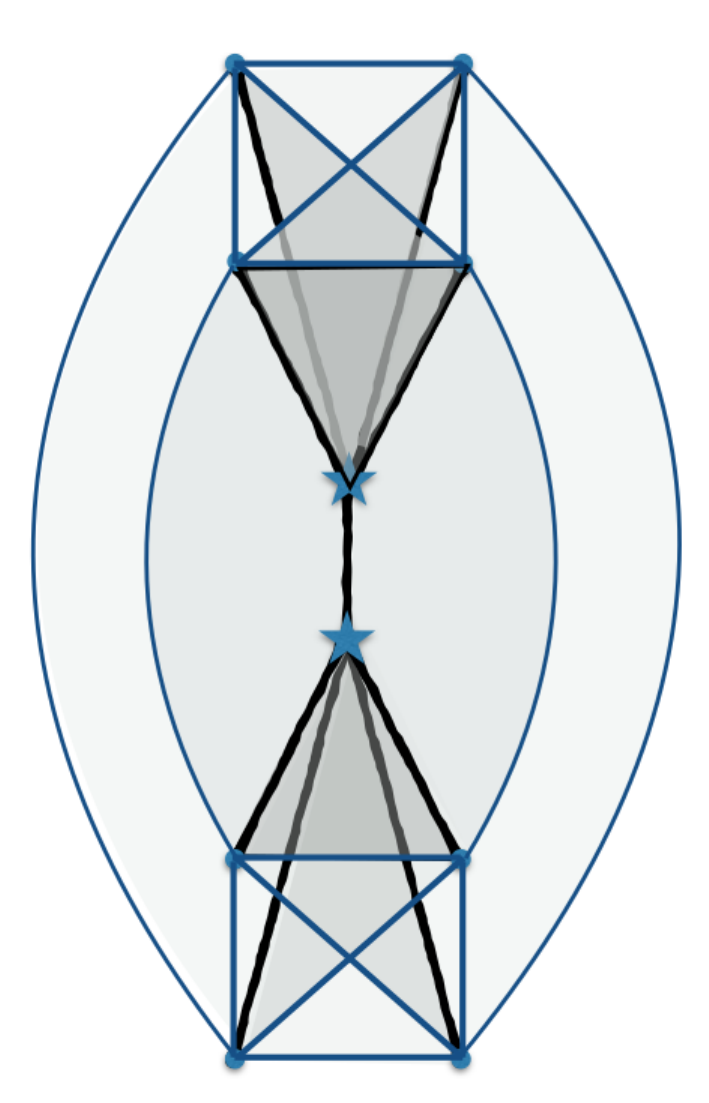} \hspace{1cm} &  
\includegraphics[scale=0.11]{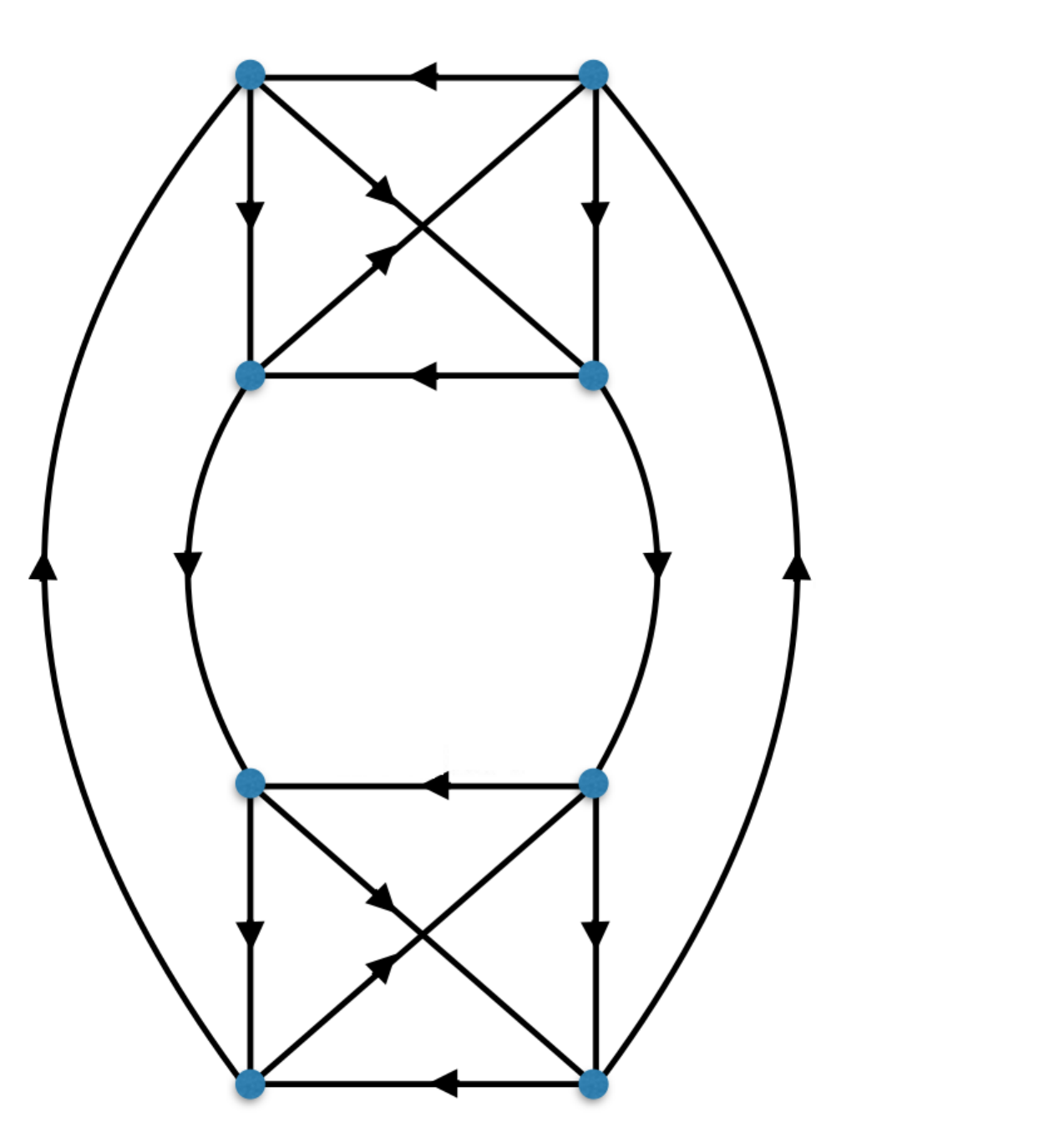} 
 \end{tabular}
\caption{The spinfoam 2--complex $\mathcal{C}$ (left) and its oriented boundary graph $\Gamma = \partial \mathcal{C}$ (right) chosen in \cite{christodoulou_planck_2016}. The four middle links (faces) carry the boundary data $\omega_\Delta$ and $\zeta_\Delta$ that correspond to a discretization of the sphere $\Delta$, defined as the intersection of $\mathcal{C^\pm}$. The six upper and six lower links (faces) carry the boundary data $\omega_\pm$ and $\zeta_\pm$ respectively, that correspond to a particularly rough discretization of the remaining of the surfaces $\mathcal{C}^\pm$ while the surfaces $\mathcal{F}^\pm$ were disregarded. It is striking that this rough discretization gives exactly the behavior for the bounce time $T_c$ and lifetime $\tau$ expected on general grounds from the analysis in Section \ref{sec:Bulk}. This should be taken as an indication that the relevant physics happen in the vicinity of the sphere $\Delta$, see \cite{mariosInfoParadox} for a detailed argument. }
\label{fig:spinfoam}
\end{figure}

The calculation done in the main text does not consider a specific choice of boundary hypersurface. In this appendix we verify numerically the estimates of Section \ref{sec:Bulk} for the crossing time $T_c$ and the lifetime $\tau$ for the specific choice of hypersurface done in \cite{christodoulou_planck_2016} (which this work completes and follows up on). The boundary data $(\omega_\ell,\vec{k}_{\ell\nn},\zeta_\ell)$ were calculated from a discretization of $\mathcal{B}$ on a 3d triangulation topologically dual to $\Gamma=\partial \mathcal{C}$. The chosen 2--complex and its boundary graph are shown in Figure \ref{fig:spinfoam}. The boundary surfaces $\mathcal{C^\pm}$ were taken to be constant Lema\^itre time surfaces and the surfaces $\mathcal{F^\pm}$ were neglected. Note that the Hessian has not been considered in the analysis that follows.

The crossing time $T_c$ is calculated from \eqref{Tc2} with the upper integration limit taken to be up to where the truncation is valid according to equation \eqref{eq:boostTruncation}. The lifetime $\tau$ is subsequently calculated from  \eqref{eq:lifetimeFormula}. The transition amplitude $W_{\mathcal{C}}(m,T)$ is approximated according to the estimate given in \eqref{eq:AmpEstimation}. The area data $\omega_\ell$ and embedding data $\zeta_\ell$ were calculated in \cite{christodoulou_planck_2016} to be  
\begin{align} \
& \omega_\Delta = 2 \left( \frac{m}{ \sqrt{2 \hbar \gamma}} \left(1+ e^{-\frac{T}{2m}}\right) \right)^2 \nonumber  \\
& \omega^\pm = 2 \left( \frac{m}{ \sqrt{12 \hbar \gamma}} \left(1+ e^{-\frac{T}{2m}}\right) \right)^2 \nonumber  \\
& \zeta_\Delta = \frac{T}{2m} \nonumber \\
& \zeta_\pm = \mp\frac{32}{9} \sqrt{6}. \label{eq:bouDataExpl}
 \end{align}
The notation for the values of the link subscript $\ell$ above is explained in the description of Figure \ref{fig:spinfoam}. These area data  completely specify the intrinsic discrete geometry at the critical point corresponding to $\omega_\ell$ and $\vec{k}_{\ell\nn}$. That is, the normal data $\vec{k}_{\ell\nn}$ can be calculated from the area data $\omega_\ell$ by basic trigonometry.

Note that the area data depend weakly on the bounce time $T$. The significant dependence on $T$ is in the embedding data $\zeta_\Delta$, that scale linearly with $T$. The data $\zeta_\Delta$ describe the scaling of the extrinsic geometry in the vicinity of the sphere $\Delta$. The embedding data $\zeta_\pm$ correspond to a smearing of the extrinsic geometry along $\mathcal{C}^\pm$ and are constant. Because the continuous surfaces $\mathcal{C}^\pm$ were chosen to be intrinsically flat, the boundary data  $\omega_\ell$ and $\vec{k}_{\ell\nn}$ determine a flat intrinsic geometry for the 3d triangulation. The last two remarks imply that this coarse discretization fails to encode the presence of strong curvature in the interior of the hole, as well as the presence of the (anti--) trapped surfaces $\mathcal{M}^\pm$. The striking result that, nevertheless, these boundary data reproduce the expected behavior for the bounce time $T_c$ and lifetime $\tau$ of Section \ref{sec:Bulk}, can be read as a strong indication that the relevant physics happens in the vicinity of $\Delta$. The reasons why this is the case are discussed in \cite{Bianchi:2018mml}. 

We find numerically that for the boundary data of equations \eqref{eq:bouDataExpl} we have
\begin{equation} \label{eq:crossingTimeInExplicit}
T_c = \frac{2 \pi}{\gamma}\, m
\end{equation}
and
\begin{equation} \label{eq:LiftimeInExplicit}
\tau \propto e^{- \frac{\Xi}{t(m)}} \; ,\quad\Xi \approx 1820.
\end{equation}
These numerical estimates are for the full expression for the amplitude estimate, as in equation \eqref{eq:AmpEstimation}. In particular, the sum over the co--frame orientation configurations $s(v)$ is included. Then, the amplitude estimate is given by the sum of four terms, corresponding to the four possible co--frame orientations for a two--vertex spinfoam. Each term in the sum is a product of sixteen gaussian weights, each corresponding to one of the sixteen faces of the spinfoam, see Figure \ref{fig:spinfoam}.

We now comment on the relevance of the fact that the boundary data in \cite{christodoulou_planck_2016} correspond to a 3d geometry. The boundary data $\omega_\ell$ and $k_{\ell\nn}$ in \cite{christodoulou_planck_2016} correspond to a critical point for the partial amplitude that reconstructs a degenerate 4d geometry. That is, two 4--simplices with triangle areas $\omega_\ell\, \hbar$ and face normals $k_{\ell\nn}$ as chosen in \cite{christodoulou_planck_2016}, and glued along one of their five  tetrahedra so that they correspond to a simplicial manifold dual to the spinfoam in Figure \ref{fig:spinfoam}, will have zero 4--volume. This can be checked explicitly by calculating the edge lengths of the 4--simplices from $\omega_\ell$ and $k_{\ell\nn}$, and then calculating their 4--volume written as a Cayley--Menger determinant, verifying that it vanishes. The vanishing of the 4--volume follows from the fact that the triangulation is taken to be intrinsically flat: the five tetrahedra making up each four simplex glue properly when embedded in a 3d Euclidean space. They correspond to a tetrahedron split in four tetrahedra with all deficit angles on the interior edges equal to zero. Thus, when promoted to a 4--simplex, this is a degenerate 4--simplex. For an analogy in one dimension lower, think of a tetrahedron with three of its triangles in the plane of the fourth triangle. This can be understood either as a 2d geometry made up of three triangles, or, as a 3d geometry made up of one tetrahedron of zero 3--volume.

 We saw in Section \ref{sec:Bulk} that the estimates for $T_c$ and $\tau$ are not affected by the kind of geometrical critical point for the partial amplitude. Then, the fact that the chosen boundary data correspond to a degenerate 4d triangulation can be seen as an (accidental) smart choice, that 
allows to understand easily equations \eqref{eq:crossingTimeInExplicit} and \eqref{eq:LiftimeInExplicit}. All dihedral angles $\phi_\ell(\delta_\ell)$ will vanish, there is only a $\Pi_\ell = \pi$ thin--wedge contribution at $\Delta$ to consider on top of the embedding data $\zeta_\ell$.  The dihedral angles $\phi(\delta_\ell)$ are calculated using well known trigonometry formulas, see for instance \cite{dittrich_area-angle_2008-1}.

Setting $\phi_\ell(\delta_\ell)=
0$ for all $\ell$ and neglecting the sum over co--frame orientations $s(v)$ and the scaling $\lambda^{2M}$ of \eqref{eq:AmpEstimation}, the transition amplitude then scales as 
\begin{equation}
W(m,T) \sim e^{- \frac{4}{t(m)} \left(\gamma \frac{T}{2m} - \pi \right)^2} e^{- \frac{12}{t(m)}  \left(\zeta^\pm\right)^2},
\end{equation}
with the factors $4$ and $12$ coming from the number of corresponding links in the boundary graph. Then, the crossing time can be read off directly from this expression as  $T_c=2 \pi m/\gamma$, in agreement with the numerical estimate in equation \eqref{eq:crossingTimeInExplicit}. Setting $T=T_c$, we have
\begin{equation}
\vert W(m,T_c) \vert^2 \sim  e^{- \frac{24}{t(m)} \, \left(\zeta^\pm\right)^2}.
\end{equation}
Thus the lifetime will scale as $\tau(m) \sim  e^{ \frac{\Xi}{t(m)} }$ with $\Xi = 24 \, (\zeta^\pm)^2 \approx 1820$, in agreement with equation  \eqref{eq:LiftimeInExplicit}.

\begin{figure}[H] 
\centering 
\includegraphics[scale=0.185,angle=0]{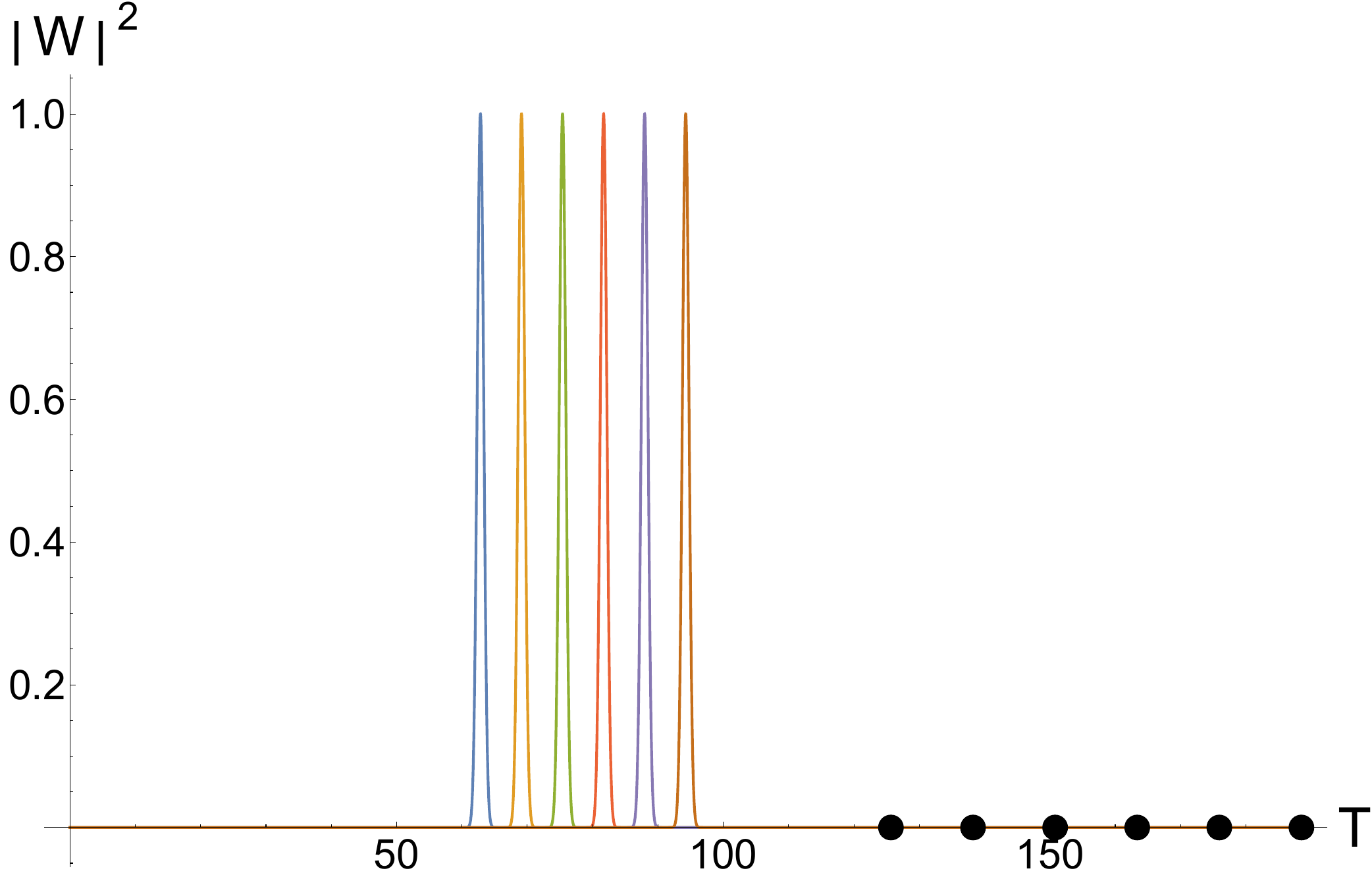} 
\caption{The modulus squared of the transition amplitude $W(m,T)$ for mass values $m=10,11,\dots,15$. The peak in the bounce time $T$ is at $T_c = 2 \pi m / \gamma $ and corresponds to the crossing time, see also Figure \ref{fig:numsImmirzi}. The peak is normalized to unit for presentation purposes. The semiclassicality parameter is fixed to $t = \hbar / m^2$ ($n=2$) and the Immirzi parameter to $\gamma=1$. The bold black dots on the horizontal axis mark the maximal value of $T$ for which the estimate for the transition amplitude of equation \eqref{eq:AmpEstimation} is valid, as a result of the truncation. According to equations \! \eqref{eq:boostTruncation} and \eqref{eq:bouDataExpl}, the estimate is valid in the interval $0\leq T \leq 4 \pi m /\gamma$. }
\label{fig:numsGauss}
\end{figure}

\begin{figure}[H] 
\centering
\includegraphics[scale=0.175,angle=0]{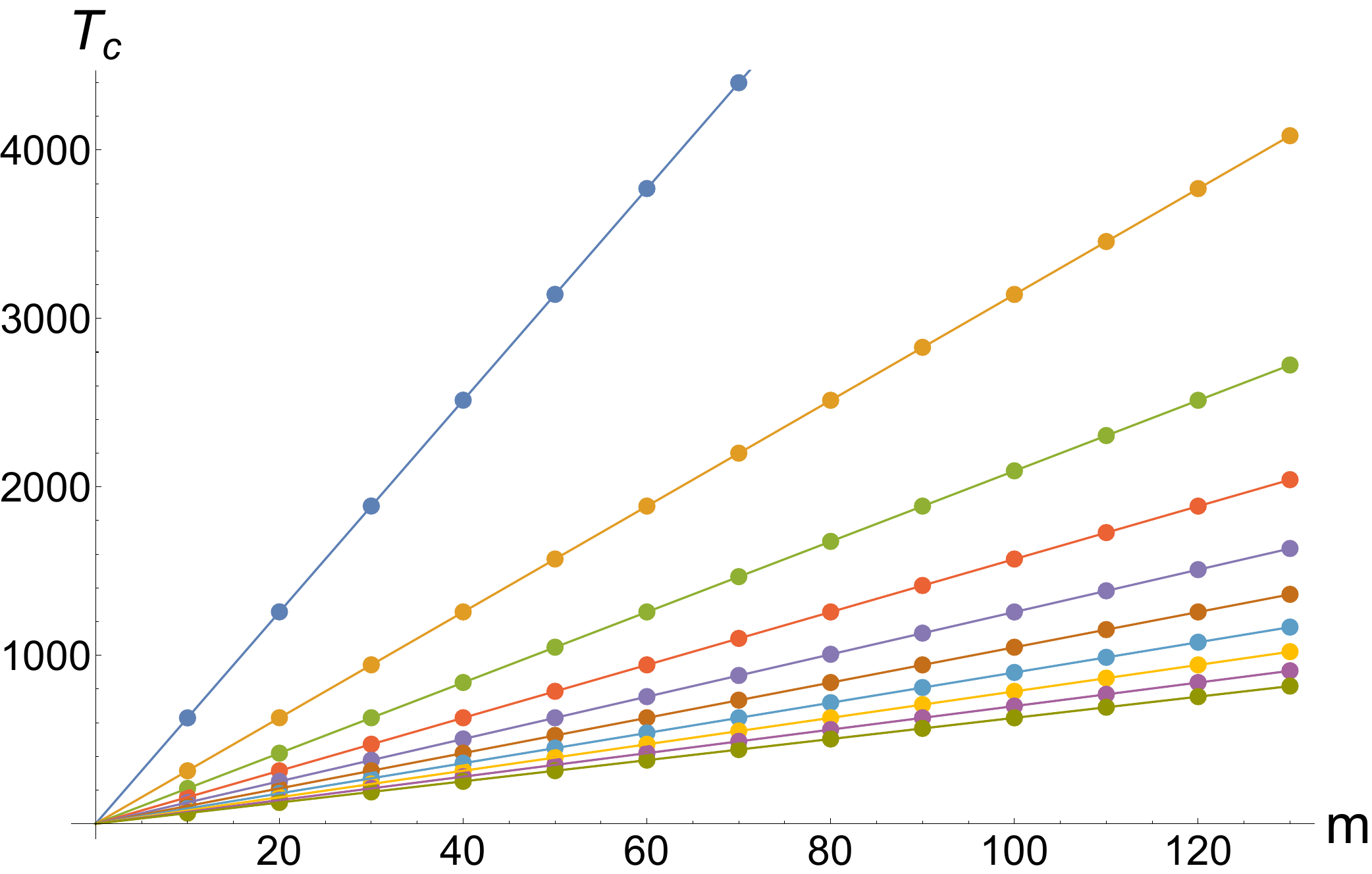} 
\caption{The crossing time $T_c$ for mass values $m=10,20,\ldots,130$ and for different values of the Immirzi parameter, $\gamma = 0.1,0.2,\ldots,1$. The interpolation is $2 \pi m /\gamma$. Numerical tests for different powers $n$ for the semiclassicality parameter $t=m^{-n}$ and different values for the Planck constant $\hbar$ give identical results, verifying that $T_c$ does not depend on $t$ and does not scale with $\hbar$.  }
\label{fig:numsImmirzi}
\end{figure}

\begin{figure}
\centering
\includegraphics[scale=0.25,angle=0]{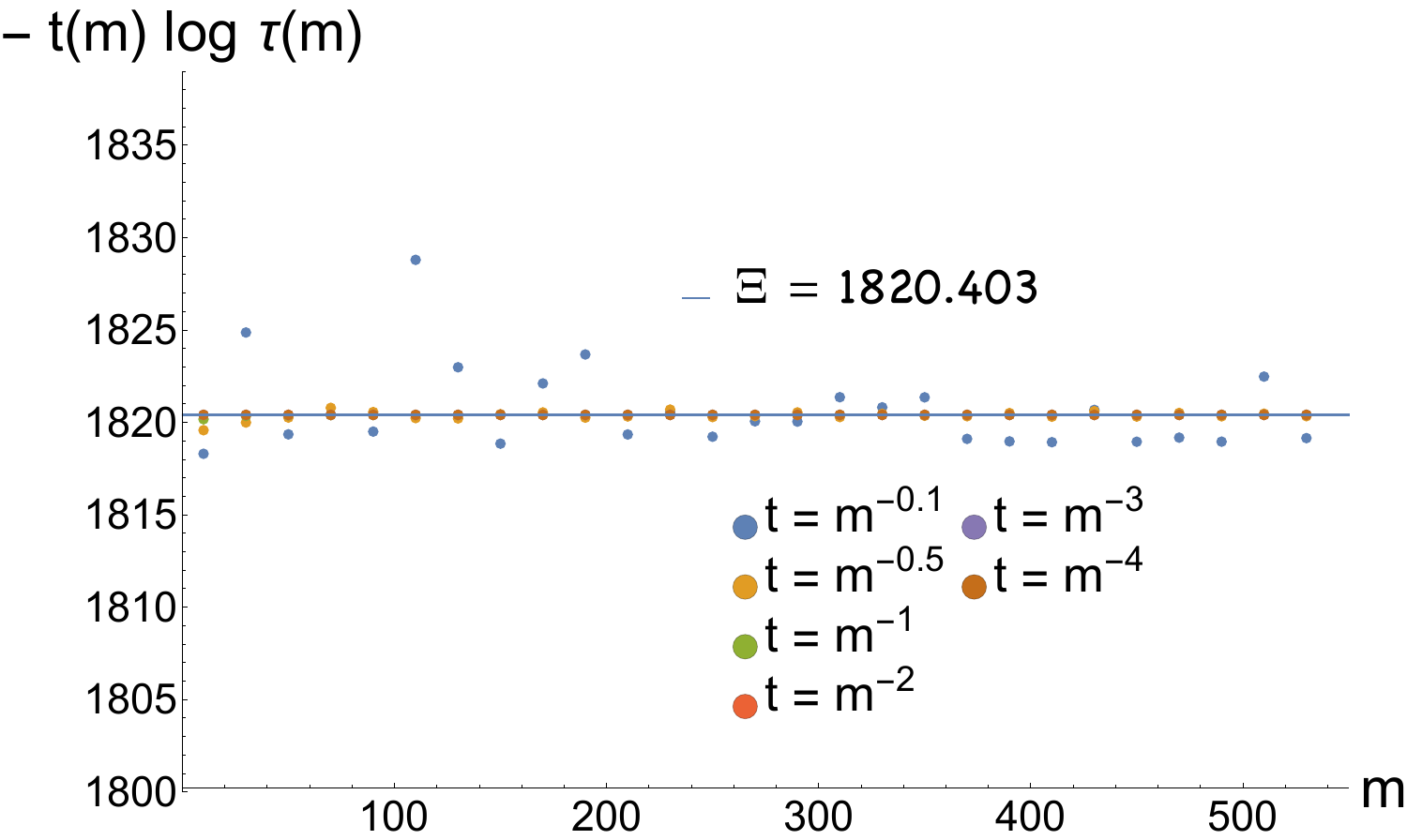} 
\caption{This and the following two figures show that the lifetime scales as $\tau(m) \sim e^{-\Xi/t(m)}$, where $\Xi$ is to a very good approximation a positive constant for the permissible values for the semiclassicality parameter $t(m)$. The estimate in eq.\! \eqref{eq:AmpEstimation} begins to break down when $n$ approaches the lower limit of eq.\! \eqref{eq:semiConstraints}. This effect is visible in the data set for $t=m^{-0.1}$ and $\hbar = 1$ (blue), which nevertheless gives $\Xi$ a constant within 1\%. The other data sets overlap within at least 0.1\% accuracy.
}
\label{fig:numsLifeConst}
\end{figure}

\begin{figure}
\centering
\includegraphics[scale=0.25,angle=0]{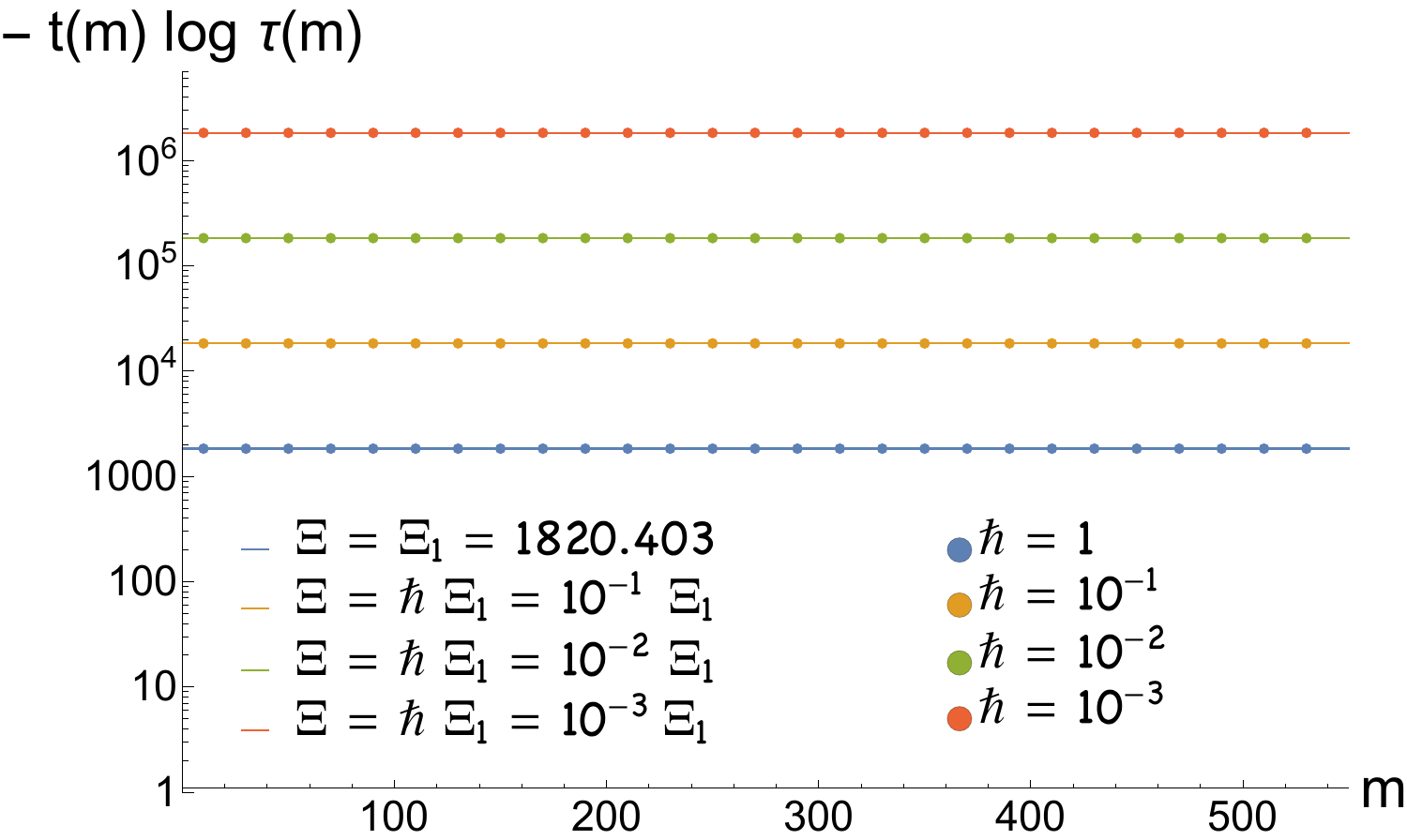} 
\caption{In this plot we verify that, as expected dimensionally, $\Xi$ scales as $\hbar^{-1}$. The vertical axis is logarithmic. The semiclassicality parameter is fixed to $t=m^{-2}$ ($n=2$) and the Immirzi parameter $\gamma$ is set to unit. The lifetime $\tau(m)$ goes to infinity as $\hbar \rightarrow 0$. Numerical tests for different values for the Immirzi parameter $\gamma$ give identical results, verifying that $\Xi$ does not depend on $\gamma$.}

\label{fig:numsLifeHbar}
\end{figure}

These results are verified numerically in the figures discussed below. We now briefly summarize their content, further details are given in the figure captions. The amplitude estimate is shown in Figure \ref{fig:numsGauss}. We see that a pronounced peak is present in the interval of the bounce time $T$ for which the estimate is reliable. The value of $T$ at the peak is the crossing time $T_c$. In Figure \ref{fig:numsImmirzi} we verify that $T_c$ is given by $T=2 \pi / \gamma$. In the following two figures we show that the lifetime scales as $\tau(m) \sim e^{-\Xi/t(m)}$ with $\Xi$ a positive constant. Instead of $\tau(m)$, we plot $-t(m) \log \tau(m)$ against $m$. In Figure \ref{fig:numsLifeConst} we see that $-t(m) \log \tau(m)$ is constant in the mass $m$ and does not depend on the power $n$.  In Figure \ref{fig:numsLifeHbar} we verify that for $t=m^2/\hbar$, $\Xi$ scales as the inverse of $\hbar$.

\section{Scaling of the Geometry in $m$ and $T$ and Monotonicity of Boost Angles} \label{sec:scalings}

In this appendix we discuss the scaling of geometrical quantities with respect to the spacetime parameters of the HR metric. In particular, we show that any boost angle $\xi$ between two timelike vectors $n_i^\alpha=(n_i^v,n_i^r,0,0)$, $i=1,2$, will scale monotonically with $X \equiv T/m$, as well as with $T$ and $m$ separately. Concretely, we find that 
\begin{equation} 
\label{eq:boostScalingRes}
\mathrm{sign} \frac{\dd \, \xi}{\dd X} =  - \mathrm{sign} f, 
\end{equation}
where $f$ is the Schwarzschild lapse function. This means that the scaling behavior is inverted when considering a boost angle calculated inside or outside the horizon, decreasing or increasing accordingly with $X$. Therefore,  the condition $\frac{\dd \xi}{\dd X} =0 $ is an equivalent characterization of the $r=2m$  hypersurfaces in the HR spacetime:
\begin{equation}
\frac{\dd \xi (r)}{\dd X} =0 \quad \Leftrightarrow \quad r=2m,
\end{equation}
where $\xi$ is any boost angle calculated at a point with coordinate radius $r$. This scaling behavior demonstrates that the embedding data $\zeta_\ell$ can encode the presence of the (anti--) trapped surfaces $\mathcal{C}^\pm$.

For definiteness, we take $n_i^\alpha$ to be both past or future oriented (thick--wedge). The case of normals with opposite time orientation (thin--wedge) proceeds similarly. See Chap.\!\! 4 of \cite{mariosPhd} for the role of the two cases in the Lorentzian Regge action. 

The boost angle $\xi$ is given by
\begin{equation} \label{eq:boostangle}
\xi = \mathrm{arcosh}\,- \frac{g(n_1,n_2)}{\vert n_1 \vert \, \vert n_2 \vert},
\end{equation}
where $\vert n_i \vert \equiv \sqrt{-g(n_i,n_i)}$ and the inner product is taken with the metric $g$. The inverse hyperbolic cosine is a real strictly monotonically increasing function when its argument is larger or equal to one, which is the case here. Specifically,
\begin{equation} \label{eq:IXrange}
I \equiv - \frac{g(n_1,n_2)}{\vert n_1 \vert \, \vert n_2 \vert} \in (1,\infty),
\end{equation}
with $I=1$ excluded because $n_1$ and $n_2$ are taken to be different vectors. Then, to conclude that boost angles scale monotonically in $X$, it suffices to show that $I$ is a monotonic function of $X$.

We want to examine the scaling of a boost angle as we move through the family of HR spacetimes, that is, as we vary $m$ and $T$. Then, the definition of the locus  at which the boost angle $\xi$ is calculated cannot depend on $m$ or $T$. The same is true for other geometrical invariants, such as proper areas etc.\! A simple way to achieve this is to use dimensionless coordinates, adapted to the spacetime parameters. As an example, consider the Schwarzschild line element in ingoing EF coordinates. Applying the coordinate transformation $r \rightarrow \tilde{r} \equiv r/m$ and $v \rightarrow \tilde{v} \equiv v/m$, we have
\begin{equation} 
\dd s^2 = m^2\left[-\left(1-\frac{2}{\tilde{r}} \right)\dd \tilde{v}^2 + 2 \dd \tilde{v} \, \dd \tilde{r}+ \tilde{r}^2 \dd \Omega^2 \right].
\end{equation}
Therefore any invariant integral taken on a submanifold of dimension $D =1,2,3,4$ will be equal to a factor $m^D = \sqrt{m^{2D}}$, coming from the square root of the induced metric, times an integral that does not depend on $m$. That is, areas scale with $m^2$, proper lengths and times with $m$ etc.
Since $m^2$ is a global conformal factor in the above line element, angles do not scale with $m$.

The same trick can be done with the HR metric, by defining
\begin{eqnarray}
\tilde{r} = \frac{r}{m} \;, \quad \tilde{v} = \frac{v}{T}.
\end{eqnarray}
 Then, the location of the shell is independent of $m$ and $T$ because
\begin{equation}
 \Theta\left(v+\frac{T}{2}\right)=\Theta\left(T\tilde{v}+\frac{T}{2}\right)=\Theta\left(\tilde{v}+\frac{1}{2}\right),
\end{equation}
and the metric \eqref{eq:downMetricNew} reads
\begin{equation} 
\dd s^2 = m^2\bigg[  - f(\tilde{r},\tilde{v}) X^2\dd \tilde{v}^2 + 2 X \dd \tilde{v} \, \dd \tilde{r} + \tilde{r}^2 \dd \Omega^2 \bigg],
\end{equation}
where we defined 
\begin{equation}
f(\tilde{r},\tilde{v}) \equiv 1- \frac{2}{\tilde{r}}\,\Theta\left(\tilde{v}+\frac{1}{2}\right). 
\end{equation}
We emphasize that the above form of the metric shows that the scaling behaviors discussed here concern the entire spacetime, they hold \emph{also} for the flat regions $I$ and $IV$. We read off for instance that areas scale as $m^2 \delta(X)$ where $\delta$ is some function of $X$. Similarly, $m^2$ is no longer a global conformal factor and angles are in general functions of $X$, scaling with both $m$ and $T$.

After these preliminary considerations  we may now show equation \eqref{eq:boostScalingRes}. The function $I(X)$
depends only on $X$ because the overall $m^2$ factor in the metric cancels in equation \eqref{eq:IXrange}. Take the point where the boost angle $\xi$ is being calculated to be given by some $\tilde{r}= \tilde{R}$, $\tilde{v}= \tilde{\cal{V}}$ and constant $\theta,\phi$. For conciseness, we denote $f\equiv f(\tilde{R},\tilde{\cal{V}})$ and define $N_1 \equiv n_1^r/n_1^v$ and $N_2 \equiv n_2^r/n_2^v$. 

Then, a few lines of algebra show that 
\begin{equation}
I(X) = \frac{F_1+F_2}{2 \sqrt{F_1 \,  F_2}}, \nonumber
\end{equation}
where the functions $F_i$ are given by
\begin{equation}
F_i =f X - 2N_i = \frac{\vert n_i \vert^2}{X (n_i^v)^2 }. 
\end{equation}
The first equality above gives $\frac{\dd F_i}{\dd X} =f$, and from the second equality we see that the functions $F_i$ are strictly positive. A simple application of the chain rule then gives
\begin{equation}
\frac{\dd I(X)}{\dd X} = \frac{f}{\sqrt{F_1 \,  F_2}} \big( 1 - I(X) \big).
\end{equation}
The term in parenthesis is strictly negative because of Eq.\! \eqref{eq:IXrange}. Thus, we have shown Eq.\! \eqref{eq:boostScalingRes}.

\section{Crossed Fingers: Mapping the HR Metric on the Kruskal Manifold} \label{sec:crossedFingers}
Here, we briefly discuss the mapping of the HR metric to the Kruskal manifold employed in \cite{haggard_quantum-gravity_2015} for the construction of the HR spacetime, which we call the ``crossed fingers'' mapping. We relate this construction to that of Section \ref{sec:RH}, and give the relation between the bounce time $T$ and the parameter $\delta$ used in \cite{haggard_quantum-gravity_2015}. The parameter $\delta$ determines where the two null shells intersect in the ``crossed fingers'' mapping.

\begin{figure} 
\centering
\includegraphics[scale=0.4]{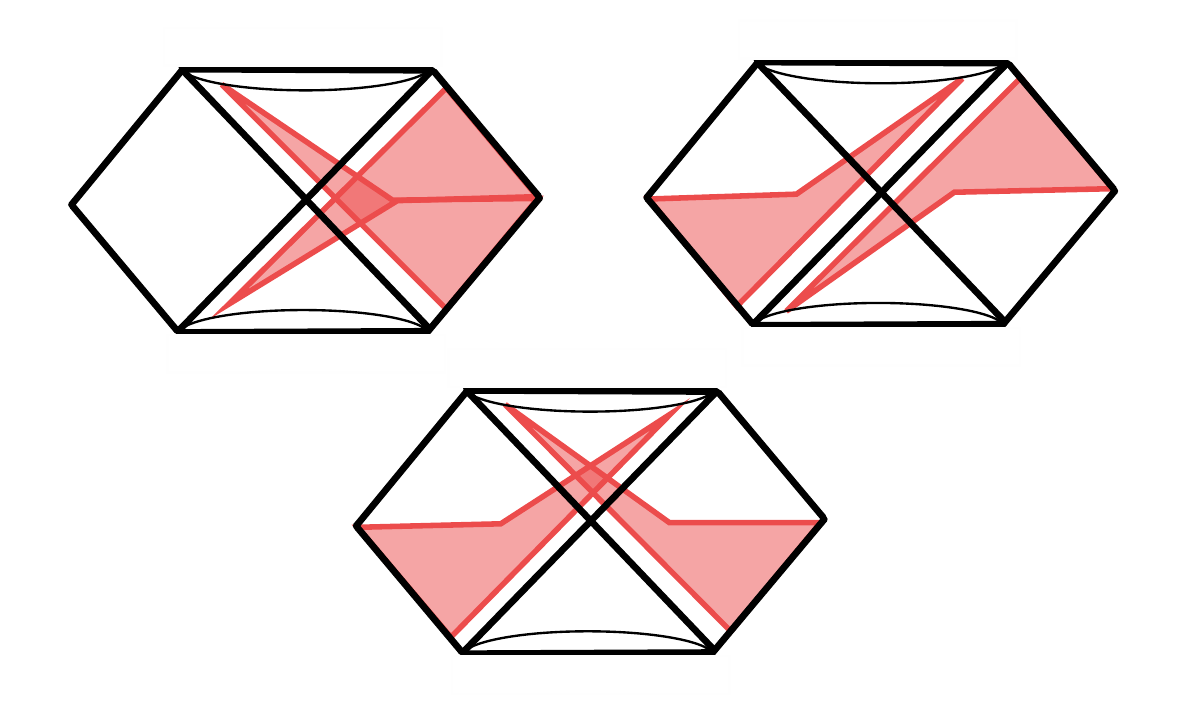}
\caption[Mappings of the HR spacetime to the Kruskal manifold.]{Some of the possible mappings of the two Kruskal patches of the HR spacetime to the full Kruskal manifold. See Figure \ref{fig:fireworksPatch} for a detailed breakdown of a single patch. It is impossible to map the HR spacetime to the Kruskal manifold using a single patch: the patches either overlap or are disjoint. Thus, we need to use at least two distinct patches. The upper--left case is the ``crossed fingers'' diagram, which corresponds to the construction originally employed in \cite{haggard_quantum-gravity_2015} and to the junction condition used here, see equation \eqref{eq:junctT}. }
\label{fig:crossedFingers}
\end{figure}

\begin{figure} 
\centering
\includegraphics[scale=0.4]{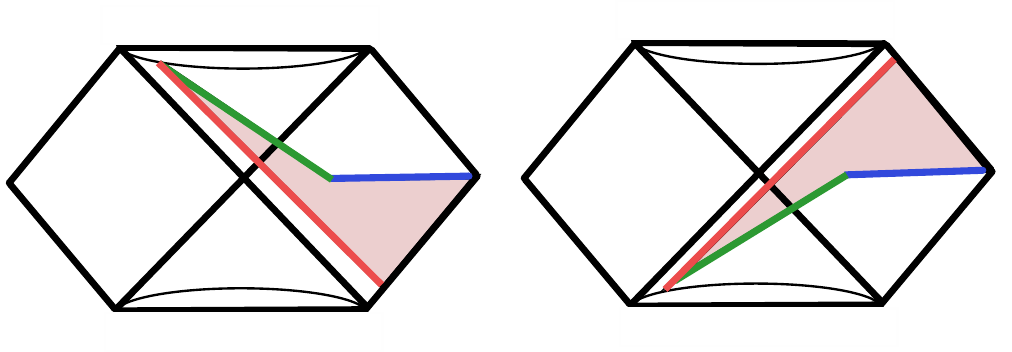}
\caption{The two Kruskal patches of the Haggard--Rovelli spacetime (color online) in the ``crossed fingers'' mapping, see top left of Figure \ref{fig:crossedFingers}. Each patch (shaded) is bounded by a null shell $\mathcal{S^\pm}$ (red), by a boundary surface $\mathcal{C^\pm}$ (green), by the fiducial surface $\mathcal{T}$ (blue) along which the two patches are joined via the junction condition of eq.\! \eqref{eq:junctT}, and by a portion of  $\cal{J}^\pm$.  The geometry of the patch on the left is given by the line elements of eqs. \eqref{eq:downMetricNew} and \eqref{eq:downMetricEF}. The geometry of the patch on the right is given by the line elements of eqs. \eqref{eq:upMetricNew} and \eqref{eq:upMetricEF}.}

\label{fig:fireworksPatch}
\end{figure}

In Section \ref{sec:RH} we described the HR metric using two different patches from the Kruskal manifold, one for region $II$ and one for region $III$ of the Carter--Penrose diagram of Figure \ref{fig:ansatz}. This is necessary because there does not exist an injective map from the union of regions $II$ and $III$ of the HR spacetime to a region of the Kruskal manifold. Different mappings are possible, all leading to the two patches either overlapping patches or being disjoint, see Figure \ref{fig:crossedFingers} and its description. The junction condition given in equation \eqref{eq:junctT} corresponds to the ``crossed fingers'' mapping, depicted on the top left of Figure \ref{fig:crossedFingers} and in more detail in Figure \ref{fig:fireworksPatch}.

We have seen that the HR metric depends on two physical scales, the mass $m$ and the bounce time $T$. The mass $m$ is implied by the use of the Kruskal manifold. The bounce time $T$, is encoded in the radius at which the two null shells cross in the ``crossed fingers'' mapping of the HR spacetime to the Kruskal manifold. We call this radius $r_\delta$ and the sphere at their intersection $\delta$. The ingoing and outgoing EF coordinates of the sphere $\delta$ are given by $v_{\mathcal{S}^-}$ and $u_{\mathcal{S}^+}$. From equation \eqref{eq:BounceTimeEF}, we infer the relation
\begin{equation}
T=-2 r^\star(r_\delta).
\end{equation}
We conclude that it is equivalent to consider the area corresponding to the radius $r_\delta$ as the second spacetime parameter for the HR metric. 

By a slight abuse of notation we introduce the parameter $\delta > 0$ for the sphere $\delta$ at radius $r_\delta$, defined by
\begin{equation}
r_\delta = 2m \left(1+\delta\right).
\end{equation}  
The bounce time $T$ and $\delta$ are then related by  
\begin{equation}
e^{-\frac{T}{4m}} = \delta \, e^{1+\delta},
\end{equation}
where we used $r^\star(r)= r+2m \log \vert \frac{r}{2m}-1 \vert$. This relation is solved for $\delta$ by the Lambert $W$ function 
\begin{equation}
\delta =W \left( \frac{e^{-\frac{T}{4m}}}{e}  \right). \
\end{equation}
The condition that the bounce time $T$ is positive translates into 
\begin{equation}
\delta < W\!\left( 1/e  \right) \approx 0.28.
\end{equation}
An infinite bounce time corresponds to a vanishing $\delta$. Thus, we may use as parameters for the HR spacetime the mass $m$, constrained to be positive, and the parameter $\delta$, constrained to lie in the interval
\begin{equation} \label{eq:deltaIntervalAllowedValues}
\delta \in \big(\,0, W\!\left(1/e  \right)\, \big).
\end{equation}

\section{The Bounce Time $T$ as an Interval at Null Infinity} \label{sec:moreT}

In this Appendix we derive the relation of our construction of the exterior spacetime to the constructions in \cite{de_lorenzo_improved_2016,bianchi_entanglement_2014}.

\subsection{The Bounce Time $T$ as an Evaporation Time and a Convenient Value for $r_\Delta$}
The bounce time $T$ can be understood as an interval of an affine parameter on $\cal{J}^+$. We will show that it is a concept analogous to the Hawking evaporation time. Despite the fact that Hawking evaporation has been neglected in this work, this alternative point of view is desirable for two reasons. First, it implies that we can directly compare time scales such as the lifetime and the crossing time, which are values for $T$, to the Hawking evaporation time scale $\sim m^3$. Second, the protrusion of the quantum region outside the trapped surfaces will interfere with the definition of the ``first'' Hawking photon. We will see that  certain constraints arise and verify that they are mild and consistent with relevant literature. 

An affine parameters on $\cal{J}^+$ is provided by the outgoing EF coordinate $u_{\mathcal{S}^+}$. From \eqref{eq:BounceTimeEF} we see that by defining an asymptotic time
\begin{equation}
u_{fhp}=v_{\mathcal{S}^-},
\end{equation}
in outgoing EF coordinates for the regions $III$ and $IV$, the bounce time corresponds to the interval 
\begin{equation}
T = u_{\mathcal{S}^+} - u_{fhp}
\end{equation}
on $\cal{J}^+$. The asymptotic time $u_{fhp}$ is light traced in the past either on the boundary surface $\mathcal{C}^+$, in which case the ray is not extendible outside region $III$, or, it will cross to region  $II$, then to region $I$ and be light traced all the way to $\cal{J}^-$. 

In the latter case, the light ray $u_{fhp}$ intersects the collapsing shell $\mathcal{S^-}$ and allows us to establish an analogy to the Hawking evaporation time. Demanding that $u_{fhp}$ is light traced to $\cal{J^-}$, is equivalent to imposing
\begin{equation} \label{eq:fhpCond}
u_\Delta \geq u_{fhp}.
\end{equation}
We will turn the above inequality into a condition for the area of the sphere $\Delta$. We first trivially  extend the outgoing EF coordinates of regions $III$ and $IV$ to region $II$ using the relation $v-u=2 r^\star(r)$ between the coordinates $(v,r)$ and $(u,r)$, and the junction condition \eqref{eq:junctT}. The new $(u,r)$ coordinate system covers the relevant portion ($u \leq u_\Delta$) of region $II$. Because the boundary $\mathcal{B}$ is arbitrary, we can only write down a necessary condition for equation \eqref{eq:fhpCond} to hold:  
\begin{equation}
r^\star(r_{\Delta}) \leq 0.
\end{equation}
This is shown as follows,
\begin{align}
 r^\star(r_{\Delta}) & \leq  0 \nonumber \\
 \Rightarrow v_\Delta - u_\Delta  & \leq   0 \leq v_\Delta - v_{\mathcal{S}^-}\nonumber \\
 \Rightarrow  \  u_\Delta  & \geq   v_{\mathcal{S}^-},
\end{align}
where we used that $v_\Delta - v_{\mathcal{S}^-} \geq 0$. 

It is convenient to define a parameter $\Delta$ by 
\begin{equation}
r_\Delta = 2 m \, (1+ \Delta) \quad ,\quad \Delta > 0.
\end{equation}
Note the abuse of notation: $\Delta$ denotes both, the sphere at the intersection of $\mathcal{C}^\pm$ and the positive number $\Delta$ related to its area by $A_\Delta  = 4 \pi \, (2m)^2 (1+\Delta)^2$. The equation $r^\star(r_{\Delta})=0$ reads
\begin{equation}
1 + \Delta + \log \Delta =0,
\end{equation}
and after exponentiation we have
\begin{equation}
 \Delta \, e^\Delta =1/e.
\end{equation}
The formal solution $\Delta_0$ to this equation is given by the Lambert $W$ function
\begin{equation}
\Delta_0 = W(1/e) \approx 0.28,
\end{equation}
and we call the corresponding radius 
$r_0$
\begin{equation}
r_0 = 2m (1+ \Delta_0) \approx 2.56 m.
\end{equation}
It follows that 
\begin{equation} \label{eq:fhpRadCond}
r_{\Delta} \leq r_0 \quad \Rightarrow \quad u_\Delta \geq u_{fhp},
\end{equation}
because $r^\star$ is monotonically increasing in $r$.

 We now consider the sphere defined as the intersection of the null hypersurface given by $u=u_{fhp}$ and the collapsing shell $\mathcal{S}^-$ which sits at $v=v_{\mathcal{S}^-}$. This sphere is in region $II$. Call the value of the radial coordinate on that sphere $r_{fhp}$. Since $u_{fhp}=v_{\mathcal{S}^-}$ and using $v-u=2 r^\star(r)$, we have
\begin{equation}
r^\star(r_{fhp}) = 0, 
\end{equation}
that is, 
\begin{equation}
r_{fhp} = r_0.
\end{equation}
The condition of eq.\! \eqref{eq:fhpRadCond} is now easy to read from a Carter--Penrose diagram. The ray $u_{fhp}$ is outgoing in the asymptotic region of the HR spacetime where the expansion of outgoing null geodesics is positive, and thus cannot intersect the sphere $\Delta$ at a greater radius than its intersection with $\mathcal{S}^-$. That is, if $r_\Delta \leq r_{fhp} = r_0 $ holds, the light ray $u_{fhp}$ can be light traced to $\mathcal{S}^-$.

The above imply that when eq.\! \eqref{eq:fhpRadCond} holds, the bounce time $T$ is analogous to an evaporation time. An evaporation time is defined as the time measured at infinity between the reception of the ``first'' and ``last'' Hawking photon. The analogue of the ``last'' Hawking photon is here the outgoing shell $\mathcal{S}^+$. A precise definition for the ``first'' Hawking photon can be found in \cite{bianchi_entanglement_2014}. In that work, the authors defined $u_{fhp}$ as marking the onset of entanglement entropy production at $\cal{J}^+$. They estimated the radius at which it is most likely for the first Hawking photon to be emitted to be roughly when the collapsing shell reaches a radius $\sim 3m$. An ambiguity of order one in the coefficient multiplying $m$ will  typically be involved in defining the emission of the first Hawking photon. 

In summary, by fixing $r_\Delta = r_0 \approx 2.58 m$, the bounce time corresponds to the interval $T = u_{\mathcal{S}^-} - u_{fhp}$ of the affine parameter $u$ at $\cal{J}^+$ and $u_{fhp}$ can be light traced to $\mathcal{S}^-$. The ray $u_{fhp}$ labels the sphere on $\mathcal{S}^-$ with radius $r_{fhp} = r_0$, a reasonable value for defining the emission of the first Hawking photon. Taking $r_\Delta = r_0$ for the extent of the quantum region does not appear restrictive. In \cite{haggard_quantum-gravity_2015} the quantum effects were estimated to be most pronounced at a radius $\frac{7}{6} 2 m \approx 2.33 m$. We see in the following section that fixing $r_0=r_\Delta$ is also convenient for other reasons.

 Of course, if Hawking evaporation is considered care must be taken for the relevant time regimes for which the metric of Section \ref{sec:RH} is a valid approximation. The discussion here makes clear that the mass loss due to Hawking evaporation can be neglected when $T \ll m^3$.

\subsection{Duration of Black and White Phases and Positivity of the Bounce Time $T$}
 Following \cite{de_lorenzo_improved_2016,bianchi_entanglement_2014}, the duration of the trapped and anti--trapped phase can be encoded in two intervals $\delta v$ and $\delta u$ respectively, defined as 
\begin{eqnarray}
\delta v & \equiv & v_\Delta - v_\mathcal{S^-} \nonumber \\
\delta u & \equiv & u_\mathcal{S^+} - u_\Delta. 
\end{eqnarray}
The meaning of $\delta v$ and $\delta u$ is as follows: For given surfaces $\mathcal{C}^-$ and $\mathcal{C}^+$, and for fixed $\delta v$ and $\delta u$, the endpoints of the portion of the $r=2m$ correspond to intervals at $\cal{J}^-$ and $\cal{J}^-$ bounded from above by $\delta v$ and $\delta u$, respectively. This can be read from Figures \ref{fig:crossedFingers} and \ref{fig:fireworksPatch}, we recall that the surfaces $\mathcal{C}^\pm$ are spacelike.

Using $v-u=2 r^\star(r)$, $\delta v$ and $\delta u$ are related to the bounce time by
\begin{equation} \label{eq:bouncetimeBHWH}
T = \delta v + \delta u - 2 r^\star(r_\Delta). 
\end{equation}
Thus, a given value for $T$ allows for different durations of the black and white hole phase. The term $-2 r^\star(r_\Delta)$ is linear in $m$ and is negligible for $T \gg m$. However, this term is negative for $r_\Delta > r_0$, and to guarantee the strict positivity of $T$ we must demand that
\begin{equation}
\delta_v + \delta_v > 2 r^\star(r_\Delta) \sim m.
\end{equation}
This is a mild condition to impose. For example, a time of order $m$ for a solar mass black hole is of the order of a microsecond, and about a second for Sagittarius $A^\star$. However, fixing $r_\Delta=r_0$ as in the previous section is again convenient. The bounce time becomes exactly the sum of $\delta v$ and $\delta u$
\begin{equation}
T = \delta v + \delta u.
\end{equation}
Since the inequalities \eqref{eq:posDeltas} ensure that $\delta v$ and $\delta u$ are always positive, $T$ is also positive
\begin{equation}
T > 0.
\end{equation}

\section{The HR metric in Kruskal Coordinates} \label{sec:KruskalRH}

In this Appendix we give the HR metric in Kruskal null coordinates, which translates our construction to the original construction appeared in \cite{haggard_quantum-gravity_2015}. We install null Kruskal coordinate systems $(U_i,V_i)$, $i = I, II, III, IV$, in all four regions of the Carter--Penrose diagram in Figure \ref{fig:ansatz}. The metric reads
\begin{equation}
\dd s^2 = -F_i(U_i, V_i) \,\dd U_i \,\dd V_i + r^2_i(U_i, V_i)\,\dd \Omega^2.\quad \nonumber
\end{equation}
In regions $I$ and $IV$ we have the flat line element
\begin{eqnarray} 
F_i(U_i, V_i) &=& 1, \nonumber \\ 
r_i(U_i, V_i) &=&\frac{V_i-U_i}{2},\nonumber
\end{eqnarray}
and in regions $II$ and $III$ the Kruskal line element
\begin{eqnarray} \label{eq:Kruskal}
F_{i}(U_{i}, V_{i}) &=& \frac{32 \,m^3}{r_{i}} \e^{-\frac{r_{i}}{2m}}, \nonumber\\ 
r_{i}(U_{i}, V_{i}) &=& 2m \, \left( W \left(-\frac{U_iV_i}{\e}\right) + 1 \right), \nonumber
\end{eqnarray}
where $W$ is the Lambert function.

The junction conditions for the intrinsic metric on $\mathcal{T}$ are trivially satisfied by identifying the coordinates of the charts in region $II$ and $III$,
\begin{eqnarray}
U_{\III} \stackrel{\mathcal{T}}{=} U_{\II}, \nonumber \\
V_{\III} \stackrel{\mathcal{T}}{=} V_{\II}.
\end{eqnarray}
 The position of the null shells $\mathcal{S^-}$ and $\mathcal{S}^+$ in these coordinates is denoted as $V_S^-$ and $U_S^+$ respectively. The junction condition for the intrinsic metric on $\mathcal{S^\pm}$ ensures that the spheres foliating these surfaces have the same area as seen by the metrics on both sides. That is, the values of the radius function on either side of $\mathcal{S^\pm}$ are identified. For $\mathcal{S^-}$, we have
\begin{equation}
 r_{\I}(U_{\I}, V_{\I} = V_{\mathcal{S^-}})= r_{\II}(U_{\II}, V_{\II} = V_{\mathcal{S^-}}). \nonumber 
\end{equation}
Equivalently,
\begin{eqnarray}
V_{\I} &\stackrel{\mathcal{S^-}}{=}& V_{\II} \stackrel{\mathcal{S^-}}{=} V_{\mathcal{S^-}} \nonumber \\ \nonumber \\
U_{\II} &\stackrel{\mathcal{S^-}}{=}& \frac{1}{V_{\mathcal{S^-}}}\left(1-\frac{V_{\mathcal{S^-}}-U_{\I}}{4m}\right) \e^{\frac{V_{\mathcal{S^-}}-U_{\I}}{4m}}. \nonumber
\end{eqnarray}
Similarly, on $\mathcal{S^+}$ we have the identification
\begin{equation}
r_{\III}(U_{\III} = U_{\mathcal{S^+}}, V_{\III})= r_{\IV}(U_{\IV} = U_{\mathcal{S^+}}, V_{\IV}), \nonumber
\end{equation} 
which gives
\begin{eqnarray}
U_{\III} &\stackrel{\mathcal{S^+}}{=}& U_{\IV} \stackrel{\mathcal{S^+}}{=} U_{\mathcal{S^+}} \nonumber \\ \nonumber \\
V_{\III} &\stackrel{\mathcal{S^+}}{=}& \frac{1}{U_{\mathcal{S^+}}}\left(1-\frac{V_{\IV}-U_{\mathcal{S^+}}}{4m}\right) \e^{\frac{V_{\IV}-U_{\mathcal{S^+}}}{4m}}. \nonumber
\end{eqnarray}
  
  For a given interior boundary, the ranges of coordinates are given by the conditions
 \begin{align}  
 &  V\!\in\!(-\infty, V_{\mathcal{S^-}}), \  U\!\in\!(-\infty, \infty), \  \ \ U\!\leq\!\mathcal{F^-}(V) \nonumber \\ 
 &  V\!\in\!(V_{\mathcal{S^-}}, \infty),\  \ \ U\!\in\!(-\infty, \infty),\ \ \ U\!\leq\!\mathcal{C^-}(V), \ \ U\!\leq\!\mathcal{T^-}(V) \nonumber \\
 & V\!\in\!(-\infty, \infty), \ \ \  U\!\in\!(-\infty, U_{\mathcal{S^+}}) , \ U\!\geq\!\mathcal{C^+}(V),\  U\!\geq\!\mathcal{T^+}(V) \nonumber \\
 & V\!\in\!(-\infty, \infty),\  \ \ U\!\in\!(U_{\mathcal{S^+}}, \infty),\ \ \ U\!\geq\!\mathcal{F^+}(V). \nonumber \\ \nonumber
\end{align}
for the patch $I,II,III,IV$ respectively. \smallskip

The coordinates of the sphere $\Delta$ must satisfy
\begin{eqnarray}
0 < V_{\mathcal{S^-}} < V_\Delta \nonumber \\
U_\Delta < U_{\mathcal{S^+}} < 0 \nonumber
\end{eqnarray} 
and, in order to ensure the presence of trapped and anti--trapped regions in the spacetime, we impose the inequalities $r_{\epsilon^\pm} < 2m$ and $r_{\Delta} > 2m$,  as in \eqref{eq:radiusRequirements}.

The transformation between Kruskal and EF coordinates on each patch is given by $V = \e^{\frac{v}{4m}}$ and $U = - \e^{\frac{-u}{4m}}$. With respect to the metric of Section \ref{sec:RH} we have 
\begin{eqnarray}
v_{\mathcal{S}^-}&=& 4m \log V_{\mathcal{S^-}}, \nonumber \\ u_{\mathcal{S}^+} &=& -4m \log - U_{\mathcal{S}^+}. \nonumber
\end{eqnarray}
 The bounce time in terms of $V_\mathcal{S}^-$ and $U_\mathcal{S}^+$ is then given by
\begin{equation}
T = 4m \log \left(- \frac{V_\mathcal{S}^-}{U_\mathcal{S}^+} \right). \nonumber
\end{equation}

\vfill

\bibliographystyle{JHEPs.bst}
\bibliography{refs.bib,newRefs.bib}

\end{document}